 \documentclass[prd,amsmath,amsfonts,preprint,11pt,superscriptaddress,onecolumn,letterpaper]{article}
\usepackage[dvips]{color}
\usepackage{amsmath}
\usepackage{amsfonts}
\usepackage{amssymb}

\usepackage{graphics}
\usepackage{amsmath}
\usepackage{amsfonts}
\usepackage{amssymb}
\usepackage{graphicx}
\usepackage{epsfig}

\def\ba{\begin{eqnarray}}
\def\ea{\end{eqnarray}}
\def\be{\begin{equation}}
\def\ee{\end{equation}}

\def\nn{\nonumber}

\def\d{\mathrm{d}}

\def\mn{_{\mu \nu}}

\def\({\left(}
\def\){\right)}

\def\ie{{\it i.e. }}
\def\eg{{\it e.g. }}

\definecolor{dunkelblau}{rgb}{0, 0, .7}
\definecolor{dunkelmagenta}{rgb}{1, 0.2, 1}
\definecolor{dunkelmagenta}{rgb}{0.95, 0.2, 0.8}
\definecolor{dunkelmagenta}{rgb}{1, 0, 0}
\definecolor{cyan}{rgb}{0,0,0.7}
\definecolor{cyand}{rgb}{0,0,0.7}

\def\crr{\color{dunkelmagenta}}
\def\cb{\color{dunkelblau}}

\title{The Effective Field Theory of Codimension-two Branes}

\author{Claudia de Rham {\thanks{e-mail address: crham@perimeterinstitute.ca}}\\[10pt]
\small{Dept. of Physics \& Astronomy, McMaster University, Hamilton ON, Canada}\\
\small{Perimeter Institute for Theoretical Physics,  Waterloo, ON, Canada}}

\begin{document}
\maketitle

\begin{abstract}
Distributional sources of matter on codimension-two and higher
branes are only well-defined as regularized objects.
Nevertheless, intuition from effective field theory suggests that the
low-energy physics on such branes should be independent of any
high-energy regularization scheme.
In this paper, we address this issue in the context of a scalar field
model where matter fields (the standard model) living on such a brane interact with
bulk fields (gravity).
The low-energy effective theory is shown to be consistent and
independent of the regularization scheme,
provided the brane couplings
are renormalized appropriately at the classical level.
We perform explicit computations of the classical renormalization group
flows at tree and one-loop level, demonstrate that the theory is
renormalizable against codimension-two divergences,
and extend the analysis
to several physical applications such as electrodynamics and
brane localized kinetic terms.
\end{abstract}

\date{}

\newpage
\tableofcontents
 \newpage

%%%%%%%%%%%%%%%%%%%%%%%%%%%%%%%%%%%%%%%%%%%%%%%
%%%%%%%%%     INTRODUCTION

%
\section{Motivations}
\label{sectionIntro}
Over the past ten years, large (supersymmetric) extra dimensions have
been subject to an increased attention, providing a simple
framework for new cosmological ideas.
Motivated by scenarios such as the Randall-Sundrum model \cite{Randall:1999ee},
physics in the presence of one extra dimension has been extensively
analyzed and represents an interesting framework in which effects
from the higher dimension can be understood.
Nevertheless, just as the dynamics of domain walls in a four-dimensional
spacetime is in many ways very different to that of a cosmic string or
point particle, the behavior of gravity near codimension-one branes is not
representative of that around higher codimensional objects.
Models with two large extra dimensions, on the other hand, are
capable of tracking some of the more interesting features of higher
codimension objects, without introducing complications
associated with higher-codimension branes.
Six-dimensional (super)gravity is therefore a
choice framework for the study of higher-dimensional effects, and
presents remarkable features of its own. Solutions of
six-dimensional gravity, have been found in
refs.~\cite{Salam:1984cj,Kanti:2001vb}, cosmological solutions in \cite{Hayakawa:2003qm}, and the stability of these models
has been studied in \cite{Graesser:2004xv,Peloso:2006cq}.
In six-dimensional supergravity, for instance, not only would the Hierarchy problem be resolved if these dimensions had
a submillimeter size \cite{Arkani-Hamed:1998rs}, but if supersymmetry remained unbroken in the
bulk at energies much lower that on the brane, the Casimir energy could also be of the
same order of magnitude as the observed four-dimensional cosmological
constant, \cite{Albrecht:2001xt}.
Another property of codimension-two objects relevant to the
cosmological constant problem is their capacity of preserving a flat
Minkowski induced geometry in the presence of any tension, \cite{Rubakov:1983bz,Aghababaie:2003wz,Burgess:2004yq,Dvali:2002pe}.

Codimension-two branes in the context of six-dimensional
(super)gravity can therefore provide potential resolutions of two of
the most embarrassing problems of current particle physics and
cosmology, namely the Hierarchy and cosmological constant problems
in scenarios where the Weinberg's argument has different imports, \cite{Weinberg:1988cp}.
Nevertheless, these models are faced with one great obstacle, the
necessity of regularizing the brane before any question can be
addressed, \cite{Geroch:1987qn}. Distributional sources of matter on codimension-two and higher
branes are indeed only well-defined as regularized objects, and one
can therefore wonder whether any regularization-independent statement
can even be made. From a field theory perspective, one expects the
low-energy theory on such a brane to be independent of any
high-energy regularization scheme, yet as of today, no
regularization-invariant scheme has been proposed to study the
effective theory on such branes. The only known work in this direction has
been developed by Goldberger and Wise in 2001 (see
ref.~\cite{Goldberger:2001tn}),
where it is pointed out that a field living on a
six-dimensional flat spacetime will typically present a pathological behavior if
coupling terms were to be introduced on a codimension-two surface.
This pathology can however be removed by appropriate renormalization
of the coupling constants. In this paper, we propose a direct
extension to ref.~\cite{Goldberger:2001tn}, where we analyze
couplings between bulk fields, free to live in the entire
six dimensions, and brane fields, which are confined to
codimension-two branes. The couplings between the two fields induce
pathologies for both fields which can be absorbed by
appropriate renormalization. Similar ideas have been proposed as being useful for understanding black hole physics
\cite{Goldberger:2004jt}, post-newtonian corrections \cite{Porto:2005ac}
and brane localized kinetic terms \cite{del Aguila:2003bh}.

In what follows, we first review in section \ref{sectionII} the problems arising when dealing
with distributional sources on codimension-two branes, the
regularization schemes that have been proposed in the literature as
well as different sources of confusion which we clarify. We then
explain the philosophy of our approach and discuss the main
consequences. Our strategy is applied in section \ref{Section free
theory}, where a scalar field toy-model is considered. In
particular, we analyze couplings between a bulk and a brane scalar
field and discuss the renormalization procedure using two
different  techniques. The first one
makes use of a conical cap to regularize the brane, while the other
removes the divergences directly in the propagators. Both methods
give rise to the same Renormalization Group (RG) flows.
 This analysis is
then extended to all possible relevant and marginal couplings in section
\ref{sectionInteractions}, where the three and four-point functions
are computed as well as the loop diagrams. Using these couplings,
we also demonstrate that the theory is
renormalizable.
The second part of this paper is then dedicated to the physical implications.
In section \ref{section em}, we show  how the same prescription
remains valid when considering the more physical example
of interactions between gravity and electromagnetism and
finally explores the implications for
localized kinetic terms on the brane which are relevant for models
such as the Dvali-Gabadadze-Porrati (DGP) in section \ref{section kinetic terms}.
We show how to make sense of
kinetic couplings on the brane and deduce that only fixed functions
of the kinetic terms are allowed by the renormalization procedure.
After concluding in section \ref{section conclusion}, we consider
all possible local counterterms in appendix \ref{appendix CT} and
argue that no such counterterms can simultaneously absorb the
bulk field divergences both in the bulk and on the brane, in
complete agreement with our procedure. Using the RG flows obtained
for the brane couplings, we finish in appendix~\ref{appendix
generalcase} by computing the arbitrary $N$-point
function at all order in loops, and prove that it remains finite in the thin-brane
limit, hence justifying that the theory is renormalizable against
codimension-two divergences.

%\part{Renormalization}
%%%%%%%%%%%%%%%%%%%%%%%%%%%%%%%%%%%%%%%%%%%%%%%
%%%%%%%%%     STRATEGY

\section{Understanding Gravity on Codimension-two branes}
\label{sectionII}
\subsection{Distributional sources}
In 1987, Geroch \& Traschen showed
that strings and point particles do not belong to the class of
metrics whose curvatures are well defined as distributions, \cite{Geroch:1987qn}. In their analysis, they considered a string in
$(3+1)$-dimensions to be regularized as a cylinder of radius $\epsilon$
carrying energy density  $\rho$. The profile of the gravitational
potential $U$ is obtained by solving Poisson's equation $\nabla^2
U=-\rho$, where $\rho$ vanishes outside the cylinder. Although
well-defined when the cylinder has a finite width $\epsilon$, it turns out that the
gravitational potential $U$ is not locally integrable in the thin-brane limit $\epsilon\rightarrow 0$
and so strings (and point particles) are not permitted as sources in
$(3+1)$-dimensions, unless they are pure tension strings.
Issues arising from smoothing out codimension-two branes are also discussed in
\cite{Corradini:2002ta}.

Despite the amount of attention branes have recently received,
mainly motivated by string theory, the situation is unfortunately no
different for those objects whose intrinsic codimension is equal or
greater to two. From the string theory point of view, progress in these areas has mainly been achieved by
neglecting the backreaction of such objects, treating them as
test particles, the so-called probe-brane approximation. From a cosmological point view, however, such
a procedure would miss some of their most important features and is
hence not always satisfying. Instead, much effort has been invested
in specific regularizations of the theory, such as
models arising from Abelian-Higgs theory, \cite{Gherghetta:2000qi},
thick brane regularizations, \cite{Navarro:2004di},
capped branes, \cite{Peloso:2006cq,Burgess:2007vi},
intersecting branes, \cite{Nelson:1999wu},
codimension-two branes confined on codimension-one objects \cite{Cod2Cod1}, etc...

In all of these examples, if the extra dimensions are compact or if an
Einstein-Hilbert term is confined on the brane, one recovers
four-dimensional gravity for the zero mode. However a logarithmic
divergence appears in the first Kaluza-Klein mode as soon as the
regularization mechanism is removed (or the thin-brane limit is
taken). Understanding the significance of this divergence and the
consequences for an observer on the brane represents the main
objective of this paper.

\subsection{Philosophy}

Bulk fields away from a defect should be insensitive to
the regularization procedure, but
evaluating them on the defect itself requires knowledge of the
internal structure of the defect.
The philosophy of this paper is therefore to accept the presence of
divergences in the thin-brane limit for bulk fields when evaluated on a codimension-two
defect but to ensure that these divergences do not propagate into
matter fields confined on the defect. We will show the validity of
an effective field theory for such fields and present
how observables on the brane remain finite after appropriate renormalization
of the coupling constants.

\subsubsection*{Origin of the problem}

Following the analysis of Goldberger and Wise
\cite{Goldberger:2001tn}, the brane-to-brane Feynman propagator for a massless scalar field
living in six-dimensional flat space-time with a conical singularity
of deficit angle $2\pi(1-\alpha)$ at $r=0$ is given by
\ba
\label{b-b prop}
D_k(0; 0)=-\int_0^{\Lambda}\frac{\d q q}{2 \pi
\alpha} \frac{i}{k^2+q^2}=-\frac{i}{2\pi\alpha} \log
\frac{\Lambda}{k}\,,
\ea
where $k$ is the four-dimensional momentum along the brane direction,
(see eq.~\eqref{Bulk prop} for more details.)
The brane-brane propagator is therefore divergent in the
thin-brane limit for which the cutoff $\Lambda$ is sent to infinity
(or for large physical scale.)
In real space, on the other hand, the free propagator is
finite outside the coincidence limit, and the
presence of the logarithmic divergence in four-dimensional momentum space is
merely a consequence to the fact that the gravitational
potential of six-dimensional gravity behaves as $x^{-3}$ in real space.
More precisely, one can express the free brane-to-brane bulk propagator in real
space as
\ba
D(0,x;0,x')&=&\int \frac{\d^3 k}{(2\pi)^3} \int_0^\infty \frac{\d q\, q}{2\pi \alpha}
\frac{e^{i \mathbf{k}\,.\,(\mathbf{x}-\mathbf{x}')}}{k^2+q^2}\nn\\
&=&\int_0^\infty \frac{\d q\, q}{2\pi \alpha}\, \frac{e^{-q \,
|\mathbf{x}-\mathbf{x}'|}}{4
\pi |\mathbf{x}-\mathbf{x}'|}=\frac{1}{8\pi^2\alpha}\,\frac{1}{
|\mathbf{x}-\mathbf{x}'|^3}\,,
\ea
where $x$ and $x'$ represent directions tangent to the codimension-two
brane and the two-point function is evaluated at $r=0$ along the
normal direction.
When the integral over the brane momentum $k$ is performed
before that over the bulk momentum $q$, the two-point function is
finite everywhere.
Nevertheless, as soon as a source is
considered at $r=0$, the logarithmic behavior of the brane-brane
two-point function in momentum space becomes relevant. This will be
seen more concretely in what follows.

\subsubsection*{General sources of confusion}
We present here two general sources of confusion that usually appear
when discovering these logarithmic divergences:
\begin{itemize}
\item The first one is related to the nature of the divergence, and
the order of magnitude at which it arises.
Any codimension-two object arises from an underlying theory (\eg
Abelian-Higgs field theory, string theory or any other underlying
theory) which will naturally provide a regularization mechanism for the brane.
We could hence argue that the notion of thin-brane limit is not physical
and one should not be concerned about any thin-brane divergences.
Yet, such a argument would be going against the principles of Effective Field Theory (EFT).
Even though the brane is expected to be regularized at some scale (\eg the string scale),
we expect from EFT that the low-energy physics is independent of the
high-energy regularization mechanism. In other words, one should not
need to understand the physics at string scale in order to
understand and make predictions about the low-energy physics.
\item Faced with this realization, one can hope that introducing
brane localized counterterms should remove the logarithmic
divergence present in \eqref{b-b prop} without effecting the bulk propagator.
Unfortunately such an approach is too naive in this context, since brane-localized counterterms will not only
affect the brane-brane propagator $D_k(0,0)$ but also the brane-bulk $D_k(r,0)$ and
bulk-bulk propagators $D_k(r,r')$ which were previously finite. Any attempt to
absorb the logarithmic divergence of the brane-brane propagator
into brane localized counterterms will then automatically result in the
introduction of further divergences.
This argument is made more
explicit in appendix \ref{appendix CT}, where we consider the most
general set of local counterterms (that remain quadratic in the scalar
field) both in the bulk and on the brane, and show explicitly that
no such counterterms will allow the propagator to be finite
everywhere. In the philosophy we will follow, we will thus not attempt to
make the bulk field propagator finite everywhere but will rather
explore the consequences for observers on the brane.
\end{itemize}

\subsubsection*{Strategy}

In a conical space-time, the propagator
diverges at the tip of the cone. The aim of this paper is to explore
the consequences for a scalar field living on the tip and coupled to
a bulk field. We expect physically that
\begin{enumerate}
\item Bulk fields evaluated away from the brane should not depend on
the regularization mechanism and thus be finite in the thin-brane
limit.
\item Bulk fields evaluated on the brane itself are
sensitive to the regularization procedure, since the position
at  which they are evaluated, \ie the position of the brane, is
dependent of the regularization. Therefore we do not require bulk
fields evaluated on the brane (at $r=0$) to be finite in the
thin-brane limit.
\item Brane fields, should have a well defined low-energy theory independent of the
brane regularization. As long as the energy scales probed by an
observer on the brane are much lower than that of the cutoff
theory, the physics we will observe should be
independent of the regularization, and hence finite in
the thin-brane limit.
\end{enumerate}
In this paper, we will follow this philosophy carefully. The case of
bulk fields with brane couplings was considered in
\cite{Goldberger:2001tn}. We here extend this analysis to matter
fields confined to the brane and draw conclusions for brane
observers.

%%%%%%%%%%%%%%%%%%%%%%%%%%%%%%%%%%%%%%%%%%%%%%%
%%%%%%%%%     TOY-MODEL WITH SCALAR FIELD

\section{Scalar Field Toy-model}
\label{Section free theory}
In this section we compute the propagator for scalar fields confined
to a codimension-two brane and coupled with a bulk scalar field.

We work in a six-dimensional flat space-time with a conical
singularity located at $r=0$:
\ba
\d s^2=\eta\mn \d x^\mu \d x^\nu+\d r^2 +r^2 \d \theta^2,
\ea
with $0\le\theta< 2\pi \alpha$, and where $2 \pi (1-\alpha)$
is the deficit angle ($\alpha \le 1$). A three-dimensional brane is located at the tip of the
cone and $x^\mu$ represents the coordinates along the brane
direction.
We use the notation $a=0,\cdots,5$; $\mu=0,\cdots,3$;\\
$y=(\theta,r)$ and $x^a=(x^\mu,y)$.

This scalar field toy-model represents a good framework for the study of effective
theories on codimension-two branes.
The theory is composed of two coupled scalar fields, namely
\begin{itemize}
\item The scalar field $\phi$ which symbolizes the bulk fields (gravity, dilaton, gauge field...)
and thus lives in six dimensions,
\item The brane field $\chi$ which symbolizes the matter fields
living on the brane (standard model) and thus confined to a
four-dimensional space-time.
\end{itemize}
The action for this system can thus be taken to be of the form
\ba
\label{FreeAction}
S=-\int
\d ^6 x
\(\frac12 (\partial_a
\phi)^2+\delta^2(y)\left[
\frac{1}{2} (\partial_\mu \chi)^2 +\frac{m^2}{2} \chi^2  +\lambda_2 \phi^2
+ \lambda \chi \phi
\right]
\),
\ea
where for simplicity we have assumed the field $\phi$ to be massless
in the bulk, but further extensions will be considered in section
\ref{sectionInteractions}.
The coupling between the bulk and brane fields is symbolized by the
term $\lambda \chi \phi$ (which can be set to zero).
Higher interactions will be considered in section
\ref{sectionInteractions}.

To make contact with previous works in the literature, we first consider a specific thick-brane regularization
mechanism, and show how the coupling constants can be renormalized
in order for the theory to remain finite in the thin-brane limit.
We then explore the renormalization
mechanism in a more systematic way by analyzing the different classical two-point
functions before turning to interactions and one-loop corrections in
the following section. We point out that both methods will give rise to the same
tree-level renomalized couplings.

\subsection{Thick-brane regularization}

As a warm up, we follow a standard technique used
in the literature to confine matter fields on a codimension-two
brane, namely a thick-brane regularization in
which the brane is no longer located at $r=0$, but rather at
$r=\epsilon$.
 \begin{figure}[h]
 \begin{center}
 \includegraphics[width=5cm]{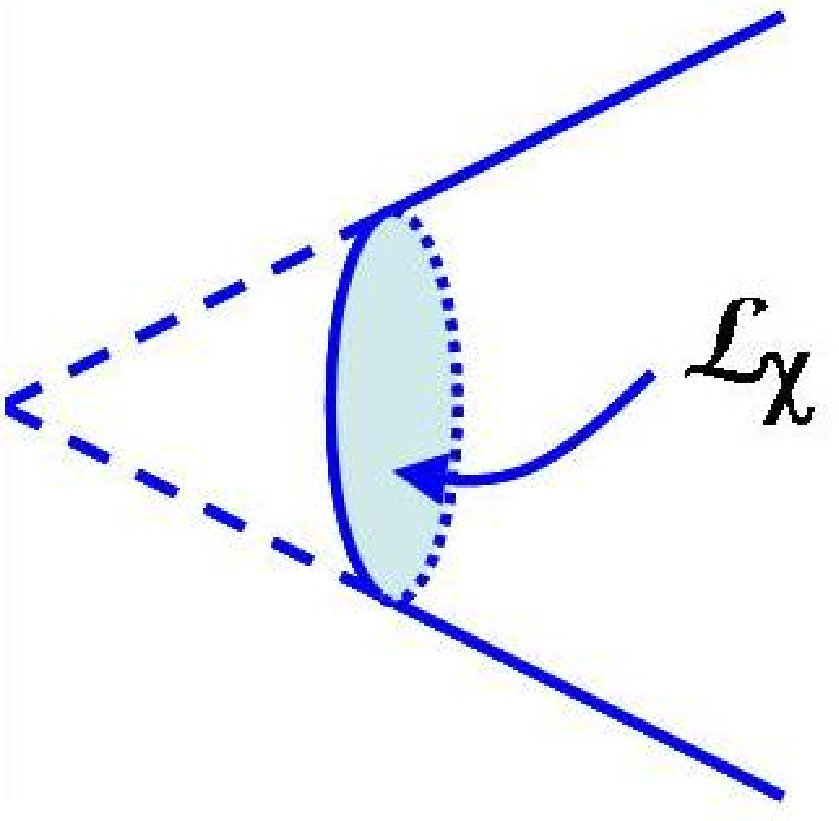}\\
 \caption{Thick brane regularization}
 \label{figure regular}
 \end{center}
 \end{figure}
The thin-brane limit is then recovered when $\epsilon \rightarrow
0$.
The action \eqref{FreeAction} is regularized by
\ba
\label{RegAction}
S=-\int \d ^4 x\d \theta \d r\, r \hspace{-8pt}&\Bigg(&\hspace{-8pt}\frac12 (\partial_r
\phi)^2
+\frac 12 (\partial_\mu
\phi)^2\\
&& \hspace{-8pt}+\frac{\delta (r-\epsilon)}{2 r \pi \alpha}\left[
\frac{1}{2} (\partial_\mu \chi)^2 +\frac {m^2}{2} \chi^2 +\lambda_2 \phi^2
+ \lambda \chi \phi
\right]
\Bigg),\nn
\ea
where for simplicity we omit for now any angular dependance.
This leads to the following equations of motions:
\ba
&&-\frac{1}{r}\partial_r(r \phi'(r))+k^2 \phi=-\frac{1}{2 \pi \alpha r}\( \lambda\chi  +\lambda_2 \phi\)\delta(r-\epsilon)\\
&& \delta(r-\epsilon)\left[(k^2+m^2)\chi=-\lambda \phi\right],
\label{bdyCond}
\ea
$k^2$ being the eigenvalue of the four-dimensional d'Alembertian
$\Box=\eta^{\mu\nu}\partial_\mu \partial_\nu$.

Integrating the first equation along the brane gives rise to the
following jump condition:
\ba
\label{Jump}
\left. r \phi'(r)\right|^{\epsilon^+}_{\epsilon^-}=-\frac{1}{2\pi
\alpha}\(\lambda\chi+\lambda_2 \phi\).
\ea

One can solve for the bulk scalar field separately in the conical
cap (for $0<r<\epsilon$) and within the bulk ($r>\epsilon$). We
choose the solution within the cap such that the scalar field
remains finite at the tip $r=0$. This leads to the following
solutions
\ba
\phi(r)=\left\{
\begin{array}{lc}
A I_0 (kr) & \text{ for } r<\epsilon\\
I_0 (kr)+B K_0 (k r) & \text{ for } r>\epsilon
\end{array}\right.\,,
\ea
where $I$ and $K$ are the two modified Bessel functions, or
hyperbolic Bessel functions,
and $I_0$ remains
finite as $r\rightarrow0$. For $r<\epsilon$, we have set the
coefficient of the divergent Bessel function $K$ to zero, so that
$\phi(r)$ remains finite as $r\rightarrow 0$. For $r>\epsilon$, on
the other hand, no such choice has been made and this solution
is therefore independent of any other boundary conditions. These results will thus stand independently to any
condition imposed on the
fields away from the brane.

The constants $A$ and $B$ are determined using the boundary
condition \eqref{bdyCond} and the jump condition \eqref{Jump}. In
the thin-brane limit, this leads to
\ba
A&=&\frac{2 \pi \alpha(k^2+m^2)}{2 \pi \alpha(k^2+m^2)- (\lambda^2-\lambda_2 (k^2+m^2))(\Gamma +\log\frac{k
\epsilon}{2})},\\
B&=&\frac{-\lambda^2+\lambda_2 (k^2+m^2)}{2 \pi \alpha(k^2+m^2)- (\lambda^2-\lambda_2 (k^2+m^2))(\Gamma +\log\frac{k
\epsilon}{2})},
\ea
where $\Gamma$ is the Euler number, $\Gamma \simeq 0.57$.

For the bulk scalar field to be well-defined in the thin-brane
limit, we require $B$ to remain finite when probing large physical scales $k\epsilon \rightarrow
0$.
This will only be possible if the logarithmic divergence is
reabsorbed into one of the coupling constants $\lambda, \lambda_2$
or $m$. Furthermore the scalar field $\chi$ should also be
well-defined in that limit. This will thus be the case if both the
following quantities remain finite
\ba
B/\chi&=&\lambda-\frac{\lambda_2}{\lambda}(k^2+m^2)\\
\chi^{-1}&=&-\frac{1}{\lambda}(k^2+m^2)+\frac{1}{2 \lambda \pi \alpha}
\(\lambda^2-\lambda_2 (k^2+m^2)\)\(\Gamma+\log\frac{k
\epsilon}{2}\).
\ea
The logarithmic divergence can thus be absorbed into the coupling
constants by arguing that their renormalized expression is related
to their bare value by
\ba
\label{RenormCouplingsThick}
\lambda_2=\frac{\lambda_{2b}}{1-\frac{\lambda_{2b}}{2 \pi \alpha}\log \rho
\epsilon} \,, \hspace{15pt}
\lambda =\frac{\lambda_{b}}{1-\frac{\lambda_{2b}}{2 \pi \alpha}\log \rho
\epsilon}\, \hspace{10pt}\text{and}\hspace{10pt}
m^2=m_b^2+\frac{\lambda^2}{\lambda_2}\,,
\ea
where the subscript $b$ represents the bare value, and $\rho$ is the physical scale.
We therefore get the following renormalization
group flows for the brane couplings
\ba
\label{RGflowthick}
\rho \frac{\d \lambda_2}{\d \rho}=\frac{\lambda_2^2}{2\pi
\alpha}\,, \hspace{15pt}
\rho \frac{\d \lambda}{\d \rho}=\frac{\lambda \lambda_2}{2\pi
\alpha}\, \hspace{10pt}\text{and}\hspace{10pt}
\rho \frac{\d m^2}{\d \rho}=\frac{\lambda^2}{2\pi
\alpha}\,.\ \
\ea
Notice that we recover the same flow as ref.~\cite{Goldberger:2001tn} for the coupling $\lambda_2$ which
gives rise to a mass term for the bulk field on the brane. The
renormalization of this coupling ensures that the bulk field to be finite
away from the tip. However we wish to emphasize that
this procedure does not get rid of the divergence of $\phi$ as $r
\rightarrow 0$. The point of the renormalization
is to make sense of the bulk field away from the tip, however at the
tip itself the bulk field diverges logarithmically as explained in
section \ref{sectionII} which is consistent with our philosophy. The
key point here is that one can still make sense of the brane field $\chi$
($\chi$ is finite) despite its coupling with $\phi$. In other words, at low-energy,
matter fields living on a codimension-two brane are independent of
the regularization procedure,
even though they couple to gravity and other bulk fields which are themselves
ill-defined in the brane in the thin-brane limit.
This is possible through adequate renormalization of the couplings.

% The existence of such renormalized couplings is quite unusual and
% could lead to interesting phenomenological results.
The renormalization for the coupling $\lambda_2$ is already known from ref.\cite{Goldberger:2001tn}.
We show here how the renormalization is extended to the couplings
for brane fields. In particular, we see that as soon as $\lambda\ne
0$, the brane field acquires a mass.

In this setup,
we have used an artificial thick brane regularization. More
fundamentally, we expect this defect to arise as the result of
other fields (\eg Abelian-Higgs scalar and gauge field $\Phi, A_\mu$), providing
a natural regularization. The fundamental theory is thus of the
form
\ba
\mathcal{P}=\int \mathcal{D}[\Phi]\mathcal{D}[A_\mu]\mathcal{D}[\phi]\mathcal{D}[\chi]
e^{i S_{\rm{tot}}[\Phi, A_\mu, \phi, \chi]}\,.
\ea
The resulting field theory \eqref{FreeAction} is obtained by integrating out the
regularizing fields $\Phi$ and $A_\mu$. In this picture, we
therefore expect that  field loop integrations generate the same
tree-level counterterms as those obtained in
\eqref{RenormCouplingsThick}.

In what follows, we shall recover the same results using a Green's
function approach, this uses the same technique as in
\cite{Goldberger:2001tn}.  We will also discuss other interaction
terms and show how the same renormalization procedures goes through,
leading to a renormalizable theory.

\subsection{EFT approach and tree-level renormalization}
In this section we adopt a more field theoretic approach and require
that the bulk-bulk propagator of the bulk field remains finite as
well as the brane field propagator. We will proceed in three steps:
\renewcommand{\labelenumi}{\roman{enumi})}
\begin{enumerate}
\item We consider first of all the purely free theory for which the
bulk and brane fields do not couple and all couplings vanish
$\lambda_2=\lambda=0$. In particular we recover the logarithmic
divergence of the bulk field propagator when evaluated on the
brane, but this divergence is only present in four-dimensional
momentum space.
\item We then consider the corrections to the bulk field propagators
arising from the mass term $\lambda_2$ on the brane. This situation
is precisely  that considered in \cite{Goldberger:2001tn}, and we
will follow the same approach. In particular, we will show how this
brane coupling induces divergences in the bulk which can be
removed by appropriate renormalization of the coupling $\lambda_2$,
hence recovering the same result as in \eqref{RenormCouplingsThick}.
\item We finally consider the corrections to both the bulk and
brane field propagators induced by the coupling $\lambda$ between
the two fields. Once again,
these couplings will induce divergences which can be removed by
renormalization of $\lambda$ and $m^2$, as in \eqref{RenormCouplingsThick}.
\end{enumerate}\vspace{15pt}

i) {\it Free propagators}\\
We concentrate first of all on the purely free theory given by the action
\ba
\label{FreeAction2}
S=-\int \d ^4 x\d \theta \d r\ r\(\frac12 (\partial_a
\phi)^2+\delta^2(y)\left[
\frac{1}{2} (\partial_\mu \chi)^2 +\frac{m^2}{2} \chi^2 \right]
\).
\ea
The propagators for both fields satisfy
\ba
\hspace{-22pt}r\Box^{(6d)}_x D(x^a, x'{}^a)\hspace{-8pt}&=&\hspace{-8pt}
\left[\partial_r(r\partial_r)+\frac{1}{r}\partial_\theta^2+r\Box_x\right]
D(x^a, x'{}^a)=i\delta^{(6)}(x^a-x'{}^a)\ \ \\
\Box_x H(x^\mu,
x'{}^\mu)\hspace{-8pt}&=&\hspace{-8pt}i\delta^{(4)}(x^\mu-x'{}^\mu)\,,
\ea
where $D$ is the Feynman propagator for $\phi$ and $H$ the one for $\chi$.
Using a mixed-representation, \ie momentum space along the
directions $x^\mu$ and real space along the two extra dimensions,
the propagators for both fields are simply
\ba
\label{Bulk prop}
D_k(r,\theta; r',\theta' )&=&-\sum_{n=-\infty}^{+\infty}\int_0^{\infty}\frac{\d q q}{2 \pi \alpha}
\frac{i}{k^2+q^2}e^{i \tilde n(\theta-\theta')}J_{|\tilde n|}(q r)J_{|\tilde n|}(q
r')\\
H_k&=&-\frac{i}{k^2+m^2}\,,
\ea
where $J_n$ the Bessel function of first kind, $\tilde n=n/\alpha$, and $k^2=\eta^{\mu \nu}k_\mu k_\nu$
the four-dimensional momentum.

Notice that in this representation, \ie in four-dimensional momentum
space, the free propagator for $\phi$ is finite when at least one of the
points is evaluated in the bulk (\ie $D_k(r,r')$ and $D_k(r,0)$
finite) but it contains a logarithmic singularity when
trying to evaluate both points on the brane.
Introducing a momentum cutoff scale $\Lambda$ in the evaluation of
the propagator, one has
\ba
\label{b-b prop 2}
D_k(0,0)=-\int_0^{\Lambda}\frac{\d q q}{2 \pi \alpha}
\frac{i}{k^2+q^2}=\frac{-i}{2\pi \alpha}\log
\frac{\Lambda}{k}\,,
\ea
which has the short distance singularity pointed out in
\cite{Goldberger:2001tn}. This divergence is usually not a
problem since the two-point function is actually finite in real
space, (see section \ref{sectionII}). However, as soon as a
source is included at $r=0$, the convolution of this two-point
function will not be finite in real space. We therefore expect this divergence to affect the two-point
function of both scalar fields when brane couplings are included.\vspace{15pt}

ii) {\it Corrections from the brane mass term $\lambda_2$}\\
The previous two-point functions were that of the free theory for
which the both fields were not coupled. We can now ``dress" these
propagators with first of all the coupling $\lambda_2$:
\ba
\label{coupledAction}
S=-\int \d ^4 x\d \theta \d r\ r\(\frac12 (\partial_a
\phi)^2+\delta^2(y)\left[
\frac{1}{2} (\partial_\mu \chi)^2 +\frac{m^2}{2} \chi^2+\frac 12 \lambda_2 \phi^2 \right]
\)\,.
\ea

\begin{figure}[h]
\begin{picture}(150,30)(-15,35)\thicklines
\large
%\linethickness{.7mm}
%
\put (0,51){\makebox(0,0){{\bf $\tilde D_k(r,r')\ =$}}}
\color{dunkelblau}
\put (40,60){\makebox(0,0){$r$}}
\put (40,50){\circle*{2}}
%\multiput(80,50)(5,0){10}{\line(1,0){3}}
\multiput(40,50)(3,0){10}{\circle*{1.3}}
\put (70,50){\circle*{2}}
\put (70,60){\makebox(0,0){$r'$}}
\color{black}
\put (85,49.5){\makebox(0,0){$=$}}
\color{dunkelblau}
\put (100,60){\makebox(0,0){$r$}}
\put (100,50){\circle*{2}}
\multiput(100,50)(10,0){3}{\line(1,0){7}}
\put (130,50){\circle*{2}}
\put (130,60){\makebox(0,0){$r'$}}
\color{black}
\put (140,49.5){\makebox(0,0){$-$}}
\color{dunkelblau}
\put (155,60){\makebox(0,0){$r$}}
\put (155,50){\circle*{2}}
\multiput(155,50)(10,0){3}{\line(1,0){7}}
\multiput(185,50)(10,0){3}{\line(1,0){7}}
\put (215,50){\circle*{2}}
\put (215,60){\makebox(0,0){$r'$}}
\put (185,60){\makebox(0,0){$0$}}
\color{cyan}
\put (185,50){\circle*{4}}
\put (215,50){\circle*{2}}
\color{black}
\put (185,40){\makebox(0,0){$\lambda_2$}}
\put (230,50){\makebox(0,0){$+$}}
\color{dunkelblau}
\put (245,60){\makebox(0,0){$r$}}
\put (245,50){\circle*{2}}
\multiput(245,50)(10,0){3}{\line(1,0){7}}
\multiput(275,50)(10,0){3}{\line(1,0){7}}
\multiput(305,50)(10,0){3}{\line(1,0){7}}
\put (335,50){\circle*{2}}
\put (335,60){\makebox(0,0){$r'$}}
\put (305,60){\makebox(0,0){$0$}}
\put (275,60){\makebox(0,0){$0$}}
\put (305,50){\circle*{4}}
\put (275,50){\circle*{4}}
\color{black}
\put (305,40){\makebox(0,0){$\lambda_2$}}
\put (275,40){\makebox(0,0){$\lambda_2$}}
\put (350,51){\makebox(0,0){$+$}}
\put (370,50){\makebox(0,0){$\cdots$}}
\normalsize
\end{picture}
\caption{Corrections to the bulk field two-point function arising from the brane mass term $\lambda_2$. The
blue dashed lines represent the free bulk field propagator
$D_k(r,r')$ while the dotted line is that corrected for the mass
term \ie $\tilde D_k(r,r')$.}
\label{FigLamdba2}
\end{figure}
%%%%%%%%%%%%%%%%%%%%%%%%%%%%%%%%%%%%

\begin{figure}[t]
\begin{picture}(150,40)(20,-5)\thicklines
\large
%\linethickness{.7mm}
%
% \put (-25,0){\makebox(0,0){{\bf $G_\phi(r,r')=$}}}
\color{dunkelblau}
\put (5,60){\makebox(0,0){$r$}}
\put (5,50){\circle*{2}}
%\put (5,0){\line(1,0){45}}
\multiput(5,50)(3,0){15}{\circle*{1.3}}
\color{cyan}
\put (27.5,50){\circle*{10}}
\color{dunkelblau}
\put (50,50){\circle*{2}}
\put (50,60){\makebox(0,0){$r'$}}
\color{black}
\put (65,49.5){\makebox(0,0){$=$}}
\color{dunkelblau}
\put (80,60){\makebox(0,0){$r$}}
\put (80,50){\circle*{2}}
\multiput(80,50)(3,0){15}{\circle*{1.3}}
\put (125,50){\circle*{2}}
\put (125,60){\makebox(0,0){$r'$}}
\color{black}
\put (140,49.5){\makebox(0,0){$+$}}
\color{dunkelblau}
\put (155,60){\makebox(0,0){$r$}}
\put (155,50){\circle*{2}}
\multiput(155,50)(3,0){10}{\circle*{1.3}}
\color{dunkelmagenta}
\put (185,49.5){\line(1,0){30}}
\color{dunkelblau}
\multiput(215,50)(3,0){10}{\circle*{1.3}}
\put (245,50){\circle*{2}}
\put (245,60){\makebox(0,0){$r'$}}
\put (185,60){\makebox(0,0){$0$}}
\put (215,60){\makebox(0,0){$0$}}
\color{cyan}
\put (185,50){\circle*{4}}
\put (215,50){\circle*{4}}
\color{cyand}
\put (185,40){\makebox(0,0){$\lambda$}}
\put (215,40){\makebox(0,0){$\lambda$}}
\color{black}
\put (260,50){\makebox(0,0){$+$}}
\color{dunkelblau}
\put (275,60){\makebox(0,0){$r$}}
\put (275,50){\circle*{2}}
\multiput(275,50)(3,0){10}{\circle*{1.3}}
\multiput(335,50)(3,0){10}{\circle*{1.3}}
\multiput(395,50)(3,0){10}{\circle*{1.3}}
\color{dunkelmagenta}
\put (305,50){\line(1,0){30}}
\put (365,50){\line(1,0){30}}
\color{dunkelblau}
\put (425,50){\circle*{2}}
\put (425,60){\makebox(0,0){$r'$}}
\put (305,60){\makebox(0,0){$0$}}
\put (335,60){\makebox(0,0){$0$}}
\put (365,60){\makebox(0,0){$0$}}
\put (395,60){\makebox(0,0){$0$}}
\color{cyan}
\put (305,50){\circle*{4}}
\put (335,50){\circle*{4}}
\put (365,50){\circle*{4}}
\put (395,50){\circle*{4}}
\color{cyand}
\put (305,40){\makebox(0,0){$\lambda$}}
\put (335,40){\makebox(0,0){$\lambda$}}
\put (365,40){\makebox(0,0){$\lambda$}}
\put (395,40){\makebox(0,0){$\lambda$}}
\color{black}
\put (440,51){\makebox(0,0){$+$}}
\put (460,50){\makebox(0,0){$\cdots$}}
%
%%%%%%%%%%%%%%%%%%%%%%%%%%%%%%%%%%%%%%%%%%%%%%%%%%%%%
%
\color{dunkelmagenta}
\put (5,0){\circle*{2}}
\put (5,0.1){\line(1,0){45}}
\put (50,0){\circle*{2}}
% \color{cyan}
\put (27.5,0){\circle*{10}}
\color{black}
\put (65,0.5){\makebox(0,0){$=$}}
\color{dunkelmagenta}
\put (80,0){\circle*{2}}
\put (80,0.1){\line(1,0){45}}
\put (125,0){\circle*{2}}
\color{black}
\put (140,0.5){\makebox(0,0){$+$}}
\color{dunkelmagenta}
\put (155,0){\circle*{2}}
\put (155,0.1){\line(1,0){30}}
\color{dunkelblau}
\put (185,10){\makebox(0,0){$0$}}
\put (230,10){\makebox(0,0){$0$}}
\multiput(185,0)(3,0){15}{\circle*{1.3}}
\color{dunkelmagenta}
\put (230,0.1){\line(1,0){30}}
\put (260,0){\circle*{2}}
\color{cyand}
\put (185,-10){\makebox(0,0){$\lambda$}}
\put (230,-10){\makebox(0,0){$\lambda$}}
\color{cyan}
\put (185,0){\circle*{4}}
\put (230,0){\circle*{4}}
\put (207.5,0){\circle*{10}}
\normalsize
\end{picture}\\
\caption{Coupling corrections to the two-point functions. The blue dotted lines represent the propagator for the bulk field $\tilde D_k(r,r')$, while the
red plain lines are the propagator for the brane field $\chi$: $H_k$.
Lines carrying a circle represent the ``dressed" propagators
$G^{\phi \phi} (r,r')$ (top diagram) and $G^{\chi \chi}$ (bottom) and take
into account the tree-level corrections arising from the coupling
$\lambda$ between the bulk and the brane field.}
\label{TreeLevel}
\end{figure}
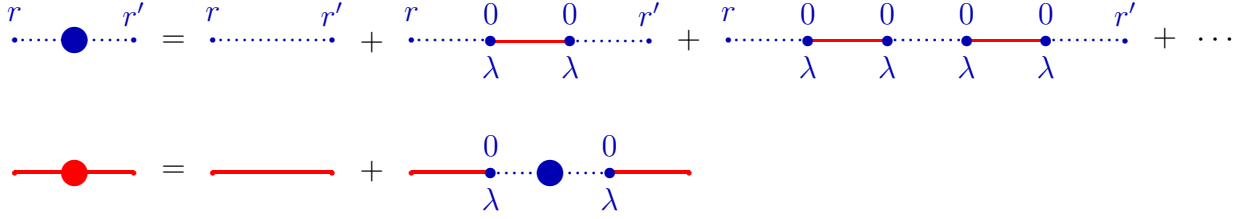

The propagator for the brane field $\chi$ remains unaffected while
that for the bulk field $\phi$ gets modified to
\ba
\tilde D_k(r,r')&=&D_k(r,r')-i\lambda_2 D_k(r,0)
D_k(0,r')+ i^2\lambda_2^2 D_k(0,0) D_k(r,0) D_k(0,r')+\cdots\nn\\
&=&D_k(r,r')-\frac{i\lambda_2}{1+i\lambda_2\,D_k(0,0)} D_k(r,0)
D_k(0,r')\,,
\ea
as symbolized in Fig. \ref{FigLamdba2}.
If $D_k(0,0)$ was finite, this bulk field propagator would be finite at
tree level as one should expect from usual field theory. In the
presented case,
the logarithmic divergence of $D_k(0,0)$ needs to be absorbed in the
coupling constant $\lambda_2$ in the following way:
\ba
\label{lambda2 renorm}
\lambda_2(\mu)=\frac{\lambda_2(\Lambda)}{1+\frac{\lambda_2(\Lambda)}{2 \pi \alpha} \log
\frac{\Lambda}{\mu}}\,,
\ea
so that this coupling constant flows as
\ba
\mu \frac{\d \lambda_2(\mu)}{\d \mu}=\frac{\lambda_2^2 (\mu)}{2\pi
\alpha}\,.
\ea
We point out that this renormalization ensures the two-point
function $\tilde D_{k}(r,r')$ to be finite away from the brane.
However, both the bulk-brane and the brane-brane two point functions
remain ill-defined: Both $\tilde D_{k}(r,0)$ and $\tilde
D_{k}(0,0)$ contains a logarithmic dependence. Once again, this is
to be expected since evaluating the bulk two-point function on the
brane requires knowledge about the exact brane position (see
section \ref{sectionII}).
\vspace{15pt}

iii) {\it Corrections from the coupling between the two fields}\\
Finally, we consider the corrections to these propagators arising
from the coupling $\lambda$ between the bulk and brane fields:
\ba
\label{coupledAction2}
S=-\int \d ^4 x\d \theta \d r\ r\(\frac12 (\partial_a
\phi)^2+\delta^2(y)\left[
\frac{1}{2} (\partial_\mu \chi)^2 +\frac{m^2}{2} \chi^2 +\frac 12 \lambda_2 \phi^2 +\lambda \phi \chi\right]
\)\,.
\ea
The tree level Green's functions for this coupled theory are symbolically represented
in Fig.~\ref{TreeLevel}.

By summing these diagrams, we obtain the
following tree-level Green's functions
 \ba
\hspace{-20pt}G^{\phi \phi}_k(r,r')
&=&\tilde D_k(r,r')-\lambda^2 \tilde D_k(r,0)\tilde D_k(0,r') H_k \sum_{n\ge0}(-\lambda^2)^n\tilde D_k(0,0)^n H_k^n\nn \\
&=& \tilde D_k(r,r')-\frac{\lambda^2 H_k}{1+\lambda^2 H_k \tilde D_k(0,0)}\tilde D_k(r,0)\tilde D_k(0,r')\nn\\
&=&D_k(r,r')-\frac{i\lambda_2+\lambda^2 H_k}{1+(i\lambda_2+\lambda^2
H_k)D_k(0,0)}D_k(r,0) D_k(0,r')\\
G^{\chi\chi}_k&=&H_k\(1- \lambda^2 G^{\phi \phi}_k(0,0) H_k\)=\frac{H_k}{1+\lambda^2 H_k
\tilde D_k(0,0)}\nn\\
&=&\frac{H_k (1+i\lambda_2 D_k(0,0))}{1+(i\lambda_2+\lambda^2
H_k)D_k(0,0)}\,.
 \ea
Notice that there is now also a mixed two-point function for the
bulk and brane fields:
\ba
G^{\phi \chi}_k(r)=\langle \phi(r)\, ,\, \chi \rangle =-\frac{i\lambda H_k D_k(r,0)}{1+\(i\lambda_2+\lambda^2 H_k\)
D_k(0,0)}\,.
\ea

Here again, if $D_k(0,0)$ was finite, both Green's functions would be finite at
the tree level as one expects in usual field theory. In the presented case,
the logarithmic divergence of $D_k(0,0)$ needs to be absorbed in the
coupling constants. These propagators will thus remain finite in the thin brane limit (for $r,r'>0$), provided the
coupling constants are renormalized as follows:
\ba
\label{RenormCouplings}
\lambda_2(\mu)=\frac{\lambda_2(\Lambda)}{1+\frac{\lambda_2(\Lambda)}{2\pi \alpha} \log
\frac{\Lambda}{\mu}}, \hspace{35pt}
\lambda(\mu)=\frac{\lambda(\Lambda)}{1+\frac{\lambda_2(\Lambda)}{2\pi \alpha} \log
\frac{\Lambda}{\mu}},\\
\label{RenormCouplingsm}
m^2(\mu)=m^2(\Lambda)-\frac{\lambda^2(\Lambda)\log\frac{\Lambda}{\mu}}{2 \pi \alpha+\lambda_2(\Lambda) \log
\frac{\Lambda}{\mu}}\,,\hspace{40pt}
\ea
leading to the following RG flows
\ba
\hspace{-10pt}\mu \frac{\d \lambda_2(\mu)}{\d \mu}=\frac{\lambda^2_2(\mu)}{2\pi \alpha}\,,\hspace{10pt}
\mu \frac{\d \lambda(\mu)}{\d \mu}=\frac{\lambda_2(\mu)\lambda(\mu)}{2\pi
\alpha}\hspace{10pt}\text{and }\hspace{5pt}
\mu \frac{\d m^2(\mu)}{\d \mu}=\frac{\lambda^2(\mu)}{2\pi \alpha}\,.
\ea
We therefore recover precisely the same relations between the bare
and renormalized coupling constants as in the thick brane analysis
\eqref{RenormCouplingsThick} and the same RG flows
\eqref{RGflowthick}. This is a non-trivial check of our
prescription.

Notice that $G^{\phi \phi}_k(r,0)$, $G^{\phi \phi}_k(0,0)$ and $G^{\phi \chi}_k(0)$ are still divergent in
the four-momentum representation, but the two-point function of the
brane field $G^{\chi\chi}_k$ has been made completely finite, and so
are $G^{\phi \phi}_k(r,r')$ and $G^{\phi \chi}_k(r)$, for $r,r'\ne0$.

In summary, we find that by renormalizing the tree-level
theory, the propagators of the field on the branes are finite, and
the propagator in the bulk are only divergent when one point is
evaluated on the brane (and in the coincident limit).
Thus there is a consistent effective field theory on the brane
and matter can be considered on a codimension-two brane in a completely
meaningful regularization-invariant way. This will have important
implications for observers on such a brane.
Before attacking this argument,
let us consider in what follows all
possible relevant and marginal interactions between a bulk and a brane field.

%%%%%%%%%%%%%%%%%%%%%%%%%%%%%%%%%%%%%%%%%%%%%%%
%%%%%%%%%     INTERACTIONS

\section{Renormalization of the relevant and marginal operators}
\label{sectionInteractions}

We consider here an extension of the precedent toy-model where
further couplings are taken into account. The effective field theory
approach will remain completely consistent after the appropriate
tree-level renormalization of the couplings. We also expect UV
divergences to be present in loop corrections, but these can be
dealt with through the usual  UV renormalization mechanism.

In order to avoid issues related to the UV divergences,
(which are independent of the fact that we consider a codimension-two brane),
we restrict ourselves to the relevant and marginal operators. The
most general brane interactions are then
\ba
S=-\hspace{-3pt}\int \d^6x\Big[
\frac 12 \(\partial \phi\)^2
+\frac{\delta(r)}{2r\pi \alpha}\,\(\frac{1}{2}\(\partial \chi\)^2+\frac
{m^2}{2}
\chi^2+\lambda \chi \phi+\frac{\lambda_2}{2} \phi^2+\mathcal{H}^{\rm{int}}_{\chi\phi}\)
\Big]\,,
\ea
with
\ba
\mathcal{H}^{\rm{int}}_{\chi\phi}=
\beta_3 \chi^3 +\beta_4
\chi^4+\lambda_3 \phi
\chi^2\,,
\ea
where the coupling $\beta_3$ is relevant while $\beta_4$ and
$\lambda_3$ are marginal.

For each diagram in this theory, we expect two sorts of divergences
to arise:
\begin{itemize}
\item The ones associated with the usual UV divergences
which appear in four dimensions when integrating over loops,
\item The short-distance divergences associated with the thin-brane
limit.
\end{itemize}
From standard four-dimensional EFT, it is a well-known fact that the interactions of the type $\beta_4 \chi^4$
will induce UV divergences in the one-loop correction of both
the two-point function and the four-point functions. These
divergences can be absorbed by renormalization of the mass $m^2$, the coupling
$\beta_4$ as well as the wave-function. However, such divergences
can be treated in a completely independent way to that arising at
the tree-level in our codimension-two scenario.
Interactions of the form $\lambda_3 \phi \chi^2$, for instance
will typically induce divergences in the thin-brane limit which
can be absorbed by appropriate renormalization of the coupling
 $\beta_3 \chi^3$, this will be studied in the three-point functions
in what follows.

\subsection{Three-point functions}

\begin{figure}[t]
\begin{picture}(150,80)(0,-40)\thicklines
\large
\color{dunkelmagenta}
\put (-10,60){\makebox(0,0){$\langle \chi\, \chi\, \chi \rangle$}}
\put (35,60){\line(1,0){30}}
\put (75,65){\line(2,1){20}}
\put (75,55){\line(2,-1){20}}
\put (50,60){\circle*{7}}
\put (85,70){\circle*{7}}
\put (85,50){\circle*{7}}
\color{black}
\put (25,60){\makebox(0,0){$=$}}
\put (110,60){\makebox(0,0){$=$}}
\color{dunkelblau}
\put (130,60){\makebox(0,0){$6\beta_3$}}
\color{black}
\put (230,60){\makebox(0,0){$+ \color{dunkelblau}{\hspace{10pt}2 \lambda_3\hspace{5pt}}\color{black}\Bigg($}}
\put (350,60){\makebox(0,0){$+ \hspace{10pt}\text{perm.}\hspace{10pt}\Bigg)$}}
\color{dunkelmagenta}
\put (145,60){\line(1,0){30}}
\put (175,60){\line(1,1){15}}
\put (175,60){\line(1,-1){15}}
\put (160,60){\circle*{7}}
\put (183,68){\circle*{7}}
\put (183,52){\circle*{7}}
\put (260,60){\line(1,0){15}}
\put (268,60){\circle*{7}}
\put (290,60){\line(1,1){15}}
\put (290,60){\line(1,-1){15}}
\put (298,68){\circle*{7}}
\put (298,52){\circle*{7}}
\color{dunkelblau}
\put (65,55){\line(0,1){10}}
\put (75,55){\line(0,1){10}}
\put (65,55){\line(1,0){10}}
\put (65,65){\line(1,0){10}}
\multiput(275,60)(3,0){6}{\circle*{1.45}}
%
%%%%%%%%%%%%%%%%%%%%%%%%%%%%%%%%%%%%%
%
\color{dunkelmagenta}
\put (-10,-10){\makebox(0,0){$\langle \, \color{dunkelblau}{\phi }\color{dunkelmagenta}\, \chi\, \chi\, \rangle$}}
\put (75,-5){\line(2,1){20}}
\put (75,-15){\line(2,-1){20}}
\put (85,0){\circle*{7}}
\put (85,-20){\circle*{7}}
\color{black}
\put (25,-10){\makebox(0,0){$=$}}
\put (110,-10){\makebox(0,0){$=$}}
\color{dunkelblau}
\multiput(145,-10)(3,0){7}{\circle*{1.45}}
\put (154,-10){\circle*{7}}
\multiput(35,-10)(3,0){10}{\circle*{1.45}}
\put (50,-10){\circle*{7}}
\multiput(330,-10)(3,0){7}{\circle*{1.45}}
\put (339,-10){\circle*{7}}
\put (130,-10){\makebox(0,0){$6\beta_3$}}
\color{black}
\put (225,-10){\makebox(0,0){$+ \color{dunkelblau}{\hspace{5pt}2 \lambda_3\hspace{5pt}}\color{black}$}}
\put (310,-10){\makebox(0,0){$+ \color{dunkelblau}{\hspace{5pt}2 \lambda_3\hspace{1pt}}\color{black}\Bigg($}}
\put (415,-10){\makebox(0,0){$+$ perm.$\Bigg)$}}
\color{dunkelmagenta}
\put (165,-10){\line(1,0){15}}
\put (180,-10){\line(1,1){15}}
\put (180,-10){\line(1,-1){15}}
\put (188,-2.5){\circle*{7}}
\put (188,-17.5){\circle*{7}}
\put (265,-10){\line(1,1){15}}
\put (265,-10){\line(1,-1){15}}
\put (273,-2.5){\circle*{7}}
\put (273,-17.5){\circle*{7}}
\put (350,-10){\line(1,0){10}}
\color{dunkelblau}\multiput(360,-9.5)(1.5,1.5){5}{\circle*{1.45}}
\color{dunkelmagenta}\put (367,-2){\line(1,1){10}}
\put (360,-10){\line(1,-1){15}}
\put (372,2.5){\circle*{7}}
\put (368,-17.5){\circle*{7}}
\color{dunkelblau}
\put (65,-15){\line(0,1){10}}
\put (75,-15){\line(0,1){10}}
\put (65,-15){\line(1,0){10}}
\put (65,-5){\line(1,0){10}}
\multiput(245,-9.5)(3,0){7}{\circle*{1.45}}
\put (254,-10){\circle*{7}}
\normalsize
\end{picture}
\caption{Three-point functions. The blue dotted lines represent the propagator for the bulk field $\phi$, while the
red plane lines are the propagator for the brane field $\chi$.}
\label{3pt}
\end{figure}

The diagrams involved in the three-point functions
are summarized in Fig.~\ref{3pt}. Summing these diagrams, we get,
for $\sum_{i=1}^3\mathbf{k}_i=0$,
\ba
G^{\chi \chi \chi}_{k_1, k_2, k_3}&=&\langle \chi_{k_1} \chi_{k_2} \chi_{k_3} \rangle\nn\\
&=&
(-i)\(6\beta_3 G^{\chi\chi}_{k_1} G^{\chi\chi}_{k_2}
G^{\chi\chi}_{k_3}+ 2\sum_{i=1}^3
\lambda_3\, G^{\phi \chi}_{k_i}(0)G^{\chi\chi}_{k_{i+1}} G^{\chi\chi}_{k_{i+2}}\)
\\
\hspace{-20pt}G^{\phi \chi \chi}_{k_1, k_2, k_3}(r)&=&
\langle \phi_{k_1}(r) \chi_{k_2} \chi_{k_3} \rangle\nn\\
&=&
(-i)\Bigg(6 \beta_3 G^{\phi
\chi}_{k_1}(r)G^{\chi \chi}_{k_2}G^{\chi \chi}_{k_3}+2\lambda_3 G^{\phi \phi}_{k_1}(r,0)
G^{\chi \chi}_{k_2}G^{\chi \chi}_{k_3}\nn\\
&&\ \ +2
\lambda_3G^{\phi\chi}_{k_1}(r)\(G^{\phi\chi}_{k_2}(0)G^{\chi \chi}_{k_3}+G^{\phi\chi}_{k_3}(0)G^{\chi
\chi}_{k_2}\)\Bigg)\,,
\ea
where the factor $(-i)$ arises from the first order expansion of
$e^{-i \int \d^4 x\, \mathcal{H}^{\rm{int}}_{\chi\phi}}$.
Notice that after appropriate renormalization of $\lambda$, $m$ and
$\lambda_2$, the propagators $G^{\chi\chi}_{k}$ and
$G^{\phi\chi}_{k}(r)$ have been made finite, however the bulk quantities
evaluated on the brane $G^{\phi \chi}_{k}(0)$ and
$G^{\phi \phi}_{k}(r,0)$ are a priori ill-defined. Once again, this divergence would propagate
into the three-point function for the brane field, had we not
renormalized the couplings $\beta_3$ and $\lambda_3$.
Upon simplification of the previous expression, we find that the
divergent part of $\langle \chi \chi \chi\rangle$  is proportional to
\ba
\langle \chi_{k_1} \chi_{k_2} \chi_{k_3}\rangle_{\rm{div}}&\propto& \(3\beta_3 -
\lambda_3 \lambda\sum_{i=1}^3\frac{iD_{k_i}(0,0)}{1+i\lambda_2
D_{k_i}(0,0)}\)\nn\,,
\ea
so that the coupling $\beta_3$ should be renormalized as
\ba
\label{beta3}
\beta_3(\Lambda)- \frac{i\lambda_3(\Lambda)\lambda(\Lambda)}{1+i\lambda_2(\Lambda) D_k(0,0)}
D_k(0,0)\hspace{40pt}\nn\\
=
\beta_3(\mu)-  \frac{\lambda_3(\mu)\lambda(\mu)}{2\pi \alpha+\lambda_2(\mu) \log \frac{\mu}{k}}\log \frac{\mu}{k}\,,
\ea
for any $k$, and
where we recall that the coupling $\lambda$ has been renormalized in such a way that
$\frac{\lambda(\Lambda)}{1+i\lambda_2(\Lambda) D_k(0,0)}$ is finite
(see eqs.~(\ref{RenormCouplings},\ref{RenormCouplingsm}).)
The divergent part of $\langle \phi(r) \chi \chi\rangle$  is then proportional to
\ba
\langle \phi_{k_1}(r) \chi_{k_2} \chi_{k_3}\rangle_{\rm{div}}&\propto& \frac{\lambda_3}{1+i\lambda_2
D_{k_1}(0,0)}\nn\,.
\ea
This divergence will thus be absorbed if the coupling $\lambda_3$ is
renormalized as
\ba
\label{RGflow3}
\lambda_3(\mu)&=&\frac{\lambda_3(\Lambda)}{1+\frac{\lambda_2(\Lambda)}{2 \pi \alpha} \log \frac \Lambda \mu}
\,,
\ea
and so
\ba
\label{beta3_2}
\beta_3(\Lambda)=\beta_3(\mu)+\frac{\lambda_3(\Lambda)\lambda(\Lambda)}
{2\pi \alpha+\lambda_2(\Lambda)\log \frac{\Lambda}{\mu}}\log
\frac{\Lambda}{\mu}\,,
\ea
Notice that this renormalization of $\lambda_3$ is precisely the one
that ensures the renormalized coupling $\beta_3$ in eq.~\eqref{beta3_2} to be
independent of the four-momentum $k$.
After renormalization, the quantity $\(\beta_3-i \lambda \lambda_3  \tilde
D_{k}(0,0)\)$ is therefore finite
\ba
\label{3ptfinite}
\mu\partial_\mu\(\beta_3-i \lambda \lambda_3  \tilde D_{k}(0,0)\)=0
\ea
which corresponds to the following flows for $\beta_3$ and $\lambda_3$
\ba
\mu \frac{\d \lambda_3(\mu)}{\d \mu}=\frac{\lambda_2(\mu)\lambda_3(\mu)}{2\pi \alpha}
\hspace{10pt}\text{and}\hspace{10pt}
\mu \frac{\d \beta_3(\mu)}{\d \mu}=\frac{\lambda(\mu)\lambda_3(\mu)}{2\pi \alpha}
\,.
\ea
Once again, this tree-level renormalization leads to a perfectly
well-defined notion of the three-point functions $G^{\chi\chi\chi}$
and $G^{\phi\chi\chi}(r)$, as long as $r$ is not evaluated on the
brane and we work outside the coincidence limit.

Furthermore, this renormalization of these coupling constants
$\beta_3$ and $\lambda_3$ also ensures that the two additional
three-point functions $\langle \phi(r_1) \phi(r_2) \chi\rangle$ and $\langle \phi(r_1) \phi(r_2)
\phi(r_3)\rangle$ are finite (for $r_i>0$). Indeed, we have,
for $\sum_{i=1}^3\mathbf{k}_i=0$,
\ba
&&\hspace{-10pt}G^{\phi\phi\chi}_{k_1,k_2,k_3}(r_1,r_2)=\langle \phi_{k_1}(r_1) \phi_{k_2}(r_2)\chi_{k_3}\rangle \nn\\
&&\hspace{10pt}=
-i\Big(6 \beta_3 G^{\phi \chi}_{k_1}(r_1) G^{\phi \chi}_{k_2}(r_2) G^{\chi
\chi}_{k_3}
+2 \lambda_3 G^{\phi \chi}_{k_1}(r_1) G^{\phi \chi}_{k_2}(r_2)
G^{\phi\chi}_{k_3}(0)\nn\\
&&\hspace{10pt}\phantom{=-i\Big(}+2 \lambda_3 \(G^{\phi
\chi}_{k_1}(r_1)
G^{\phi
\phi}_{k_2}(r_2,0)+(1\leftrightarrow
2)\) G^{\chi \chi}_{k_3}\Big)\nn\\
&&\hspace{10pt}=
-2i  G^{\chi \chi}_{k_3}\prod_{i=1}^2 G^{\phi \chi}_{k_i}(r_i)\Bigg(3 \beta_3(\Lambda)
-\lambda_3(\Lambda) \frac{i\lambda(\Lambda) D_{k_3}(0,0)}{1+i\lambda_2 (\Lambda) D_{k_3}(0,0)}-\sum_{i=1}^2\frac{\lambda_3(\Lambda)}{i\lambda(\Lambda)
H_{k_i}(\Lambda)}\Bigg)\nn\\
&&\hspace{10pt}=-2i  G^{\chi \chi}_{k_3}\prod_{i=1}^2 G^{\phi
\chi}_{k_i}(r_i)
\Bigg(
3\beta_3(\mu)-\lambda_3(\mu)\sum_{i=1}^3 \frac{\lambda(\mu)}{1+\frac{\lambda_2(\mu)}{2\pi \alpha}\log \frac{\mu}{k}}\frac{1}{2\pi \alpha}\log
\frac{\mu}{k}\nn\\
&&\hspace{10pt}\phantom{=-2i  G^{\chi \chi}_{k_3}\prod_{i=1}^2
G^{\phi
\chi}_{k_i}(r_i)
\Bigg(}+\sum_{i=1}^2\frac{\lambda_3(\mu)}{1+\frac{\lambda_2(\mu)}{2\pi
\alpha}\log
\frac{\mu}{k}}\frac{D_{k_i}(r_i,0)}{G^{\phi\chi}_{k_i}(r_i)}\Bigg)\nn\,,
\ea
which, in terms of the renormalized coupling constants, is clearly
finite. The last three-point function can be expressed in a similar
way:
\ba
&&G^{\phi\phi\phi}_{k_1,k_2,k_3}(r_1,r_2,r_3)=\langle \phi_{k_1}(r_1) \phi_{k_2}(r_2)\phi_{k_3}(r_3)\rangle \nn\\
&&\hspace{20pt}=
-i\(\prod_{i=1}^3G^{\phi \chi}_{k_i}(r_i)\)\Big(6 \beta_3
+2 \lambda_3 \sum_{i=1}^3  \frac{G^{\phi\phi}_{k_i}(r_i,0)}{G^{\phi\chi}_{k_i}(r_i)}\Big)\nn\\
&&\hspace{20pt}=
-2i \(\prod_{i=1}^3G^{\phi \chi}_{k_i}(r_i)\)
\sum_{i=1}^3\Bigg[ \beta_3(\mu)-\lambda_3(\mu) \frac{\lambda(\mu)}{1+\frac{\lambda_2 (\mu)}{2\pi \alpha} \log \frac \mu
k_i}\frac{1}{2\pi \alpha} \log \frac \mu k\nn\\
&&\hspace{150pt}+\frac{\lambda_3(\mu)}{1+\frac{\lambda_2 (\mu)}{2\pi \alpha} \log \frac \mu
k_i} \frac{D_{k_i}(r_i,0)}{G^{\phi\chi}_{k_i}(r_i)}
\Bigg]\nn\,,
\ea
and is therefore also completely finite. This result is already non-trivial as
it stands, since all four three-point functions have been made finite
by simple tree-level renormalization of the two coupling constants $\beta_3$
and $\lambda_3$. We can however push this analysis even a step further
by studying the implications for the four-point functions as well as
the loop corrections to the two-point functions.

\subsection{Four-point functions}
The classical contributions to the four-point function $\langle \chi^4 \rangle$ are
symbolically represented in Fig.~\ref{4pt}.

Summing these diagrams, we get for $\sum_{i=1}^4 \mathbf{k_i}=0$,
\ba
&&\hspace{-10pt}G^{\chi\chi\chi\chi}_{k_1,k_2,k_3,k_3}=
\langle \chi_{k_1} \chi_{k_2}\chi_{k_3}\chi_{k_4}\rangle\nn\\
&&\hspace{0pt}=\(\prod_{i=1}^4 G^{\chi \chi}_{k_i}\)\Bigg[(-i) 4!
\beta_4
+\frac{(-i)^2}{2!}\sum_{\rm{perm}_u}
2\Bigg\{6^2 \beta_3^2 G^{\chi \chi}_{k_u}+2 . 12 \beta_3 \lambda_3
G^{\phi\chi}_{k_u}(0)\hspace{25pt}\\
&&+12 G^{\chi\chi}_{k_u}\sum_{i=1}^4
\frac{G^{\phi\chi}_{k_i}(0)}{G^{\chi\chi}_{k_i}}
+4\lambda_3^2 G^{\phi\phi}_{k_u}(0,0)+4\lambda_3^2G^{\phi\chi}_{k_u}(0)\sum_{i=1}^4
\frac{G^{\phi\chi}_{k_i}(0)}{G^{\chi\chi}_{k_i}}\nn\\
&&+4\lambda_3^2G^{\chi\chi}_{k_u}\sum_{i=u_1,u_2}\sum_{j=u_3,u_4}\frac{G^{\phi\chi}_{k_i}(0)}{G^{\chi\chi}_{k_i}}
\frac{G^{\phi\chi}_{k_j}(0)}{G^{\chi\chi}_{k_j}}\Bigg\}
\Bigg]\,,\nn
\ea
where we recall again that the factor $(-i)$ and $(-i)^2/2!$ arise
from the expansion to first order (resp. second order) of
$e^{-i \int \d^4 x\, \mathcal{H}^{\rm{int}}_{\chi\phi}}$.
The sum over perm$_u$ is the one over the three permutations
$u=\{(1234),\, (1324)\,,(1423)\}$, for which $k_u=k_{u_1}+k_{u_2}=\{k_1+k_2,\,
k_1+k_3,\,k_1+k_4\}$.

\begin{figure}[t]
\begin{picture}(150,90)(-10,-65)\thicklines
%\large
\crr
\put (-20,60){\makebox(0,0)[l]{$\langle \chi^4 \rangle
\, \color{black}{=}$}}
\put (-20,-40){\makebox(0,0)[l]{$\phantom{\langle \chi^4 \rangle}
\, \color{black}{=}$}}
\crr
\put (60,60){\line(-1,1){10}}
\put (60,60){\line(-1,-1){10}}
\put (60,60){\line(1,1){10}}
\put (60,60){\line(1,-1){10}}
\put (55,55){\circle*{4}}
\put (65,55){\circle*{4}}
\put (55,65){\circle*{4}}
\put (65,65){\circle*{4}}
\put (165,60){\line(-1,1){10}}
\put (160,65){\circle*{4}}
\put (165,60){\line(-1,-1){10}}
\put (160,55){\circle*{4}}
\put (165,60){\line(1,0){15}}
\put (172,60){\circle*{4}}
\put (180,60){\line(1,1){10}}
\put (185,65){\circle*{4}}
\put (180,60){\line(1,-1){10}}
\put (185,55){\circle*{4}}
\put (277,60){\line(-1,1){10}}
\put (277,60){\line(-1,-1){10}}
\cb\multiput(277,60)(2,0){7}{\circle*{1.45}}\crr
\put (290,60){\line(1,0){10}}
\put (300,60){\line(1,1){10}}
\put (300,60){\line(1,-1){10}}
\put (272,55){\circle*{4}}
\put (272,65){\circle*{4}}
\put (295,60){\circle*{4}}
\put (305,55){\circle*{4}}
\put (305,65){\circle*{4}}
\put (344,66){\line(-1,1){10}}
\put (350,60){\line(-1,-1){10}}
\put (350,60){\line(1,0){10}}
\put (360,60){\line(1,1){10}}
\put (360,60){\line(1,-1){10}}
\cb\multiput(350,60)(-1,1){7}{\circle*{1.45}}\crr
\put (345,55){\circle*{4}}
\put (340,70){\circle*{4}}
\put (355,60){\circle*{4}}
\put (365,55){\circle*{4}}
\put (365,65){\circle*{4}}
\color{white}
\put (334,76){\circle*{4.5}}
\crr
\put (190,10){\line(-1,1){10}}
\put (190,10){\line(-1,-1){10}}
\cb\multiput(190,10)(2,0){6}{\circle*{1.45}}\put (195,10){\circle*{4}}\crr
\put (200,10){\line(1,1){10}}
\put (200,10){\line(1,-1){10}}
\put (185,15){\circle*{4}}\put (185,5){\circle*{4}}\put (205,15){\circle*{4}}\put (205,5){\circle*{4}}
\put (249,16){\line(-1,1){10}}
\put (255,10){\line(-1,-1){10}}
\put (255,10){\line(1,0){10}}
\put (265,10){\line(1,-1){10}}
\put (271,16){\line(1,1){10}}
\cb\multiput(255,10)(-1,1){7}{\circle*{1.45}}\crr
\cb\multiput(265,10)(1,1){7}{\circle*{1.45}}\crr
\put (250,5){\circle*{4}}
\put (245,20){\circle*{4}}
\put (260,10){\circle*{4}}
\put (270,5){\circle*{4}}
\put (275,20){\circle*{4}}
\color{white}
\put (239,26){\circle*{4.5}}
\put (281,26){\circle*{4.5}}
\crr
\put (322,10){\line(-1,1){10}}
\put (322,10){\line(-1,-1){10}}
\cb\multiput(322,10)(2,0){7}{\circle*{1.45}}\crr
\put (335,10){\line(1,0){10}}
\put (351,16){\line(1,1){10}}
\put (345,10){\line(1,-1){10}}
\put (317,5){\circle*{4}}
\put (317,15){\circle*{4}}
\put (340,10){\circle*{4}}
\put (350,5){\circle*{4}}
\put (355,20){\circle*{4}}
\cb\multiput(345,10)(1,1){7}{\circle*{1.45}}\crr
\color{white}
\put (361,26){\circle*{4.5}}
\crr
\color{black}
\put (30,60){\makebox(0,0){ $ -4! i \ \crr \beta_4$}}
\footnotesize
\put (87,50){\makebox(0,0)[l]{ $3 \rm{perm.}$}}
\normalsize
\put (70,60){\makebox(0,0)[l]{ $ -\frac12\, \ \sum\   2\Bigg[ $}}
\put (133,60){\makebox(0,0){ $\ 6^2\, \crr \beta_3^2 $}}
\put (230,60){\makebox(0,0){$+\ 2 . 1\!2\ \crr \beta_3 \cb \lambda_3 \color{black}\Bigg($}}
\put (320,60){\makebox(0,0){$+\, 2$}}
\put (385,60){\makebox(0,0){$\Bigg)$} }
\put (130,10) {\makebox(0,0)[l]{$+ \, 4\, \cb \lambda_3^2\,  \color{black}\Bigg($}}
\put (220,10){\makebox(0,0)[l]{$+\,4$}}
\put (290,10){\makebox(0,0)[l]{$+\,4$}}
\put (370,10){\makebox(0,0)[l]{$\Bigg) \Bigg]$}}
%
%%%%%%%%%%%%%%%%%%%%%%%%%%%%
%
%
\crr
\put (60,-40){\line(-1,1){10}}
\put (60,-40){\line(-1,-1){10}}
\put (60,-40){\line(1,1){10}}
\put (60,-40){\line(1,-1){10}}
\put (55,-45){\circle*{4}}
\put (65,-45){\circle*{4}}
\put (55,-35){\circle*{4}}
\put (65,-35){\circle*{4}}
\put (140,-35){\line(-1,1){10}}
\put (140,-45){\line(-1,-1){10}}
\put (175,-35){\line(1,1){10}}
\put (175,-45){\line(1,-1){10}}
\put (150,-40){\line(1,0){15}}
\cb \put (140,-35){\line(1,0){10}}\put (140,-45){\line(1,0){10}}\put (140,-35){\line(0,-1){10}}\put (150,-35){\line(0,-1){10}}
    \put (165,-35){\line(1,0){10}}\put (165,-45){\line(1,0){10}}\put (165,-35){\line(0,-1){10}}\put (175,-35){\line(0,-1){10}}
\crr
\put (135,-30){\circle*{4}}
\put (135,-50){\circle*{4}}
\put (180,-30){\circle*{4}}
\put (180,-50){\circle*{4}}
\put (157,-40){\circle*{4}}
\put (240,-40){\line(-1,1){10}}
\put (240,-40){\line(-1,-1){10}}
\cb\multiput(240,-40)(2,0){6}{\circle*{1.45}}\crr
\put (250,-40){\line(1,1){10}}
\put (250,-40){\line(1,-1){10}}
\put (235,-35){\circle*{4}}\put (235,-45){\circle*{4}}\put (255,-35){\circle*{4}}\put (255,-45){\circle*{4}}
\color{black}
\put (30,-40){\makebox(0,0){ $ -4! i \ \crr \beta_4$}}
\footnotesize
\put (87,-50){\makebox(0,0)[l]{ $3 \rm{perm.}$}}
\normalsize
\put (70,-40){\makebox(0,0)[l]{ $ -\phantom{\frac12}\, \ \sum\ \  \Bigg[ $}}
\put (190,-40) {\makebox(0,0)[l]{$+ \, 4\, \cb \lambda_3^2$}}
\put (270,-40) {\makebox(0,0)[l]{$\Bigg]$}}
\end{picture}
\caption{Classical contributions to the four-point functions. The second term on the bottom line is finite, but the third diagram
involves a divergent piece proportional to $\lambda_3^2 \tilde D_k
(0,0)$ which needs to be absorbed into the coupling $\beta_4$.}
\label{4pt}
\end{figure}

We now recall that the product of the two
three-point functions $\langle \chi_{k_1}\chi_{k_2}\chi_{k_u}\rangle\langle
\chi_{k_3}\chi_{k_4}\chi_{k_u}\rangle$ is {\it finite}
(after appropriate renormalization of the couplings $\beta_3$ and $\lambda_3$ in \eqref{RGflow3})
and is given by
\ba
&&\hspace{-20pt}\mathcal{A}_{u}= \langle
\chi_{k_{u_1}}\chi_{k_{u_2}}\chi_{k_u}\rangle\langle
\chi_{k_{u_3}}\chi_{k_{u_4}}\chi_{k_u}\rangle\nn\\
&&\hspace{-10pt}=-\(G^{\chi\chi}_{k_u}\)^2\(\prod_{i=1}^4G^{\chi
\chi}_{k_i}\)\(6 \beta_3
+2\lambda_3\sum_{i={u_1,u_2,u}} \frac{G^{\phi \chi}_{k_i}(0)}{G^{\chi \chi}_{k_i}}\)
\(6 \beta_3
+2\lambda_3\sum_{j={u_3,u_4,u}} \frac{G^{\phi \chi}_{k_j}(0)}{G^{\chi \chi}_{k_j}}\)\nn\\
&&\hspace{-10pt}=-\Bigg(6^2 \beta_3^2 G^{\chi\chi}_{k_u}+12 \beta_3
\lambda_3
G^{\chi\chi}_{k_u} \sum_{i=1}^4 \frac{G^{\phi\chi}_{k_i}(0)}{G^{\chi\chi}_{k_i}}
+24 \beta_3 \lambda_3 G^{\phi\chi}_{k_u}(0)
+4\lambda_3^2 G^{\phi\chi}_{k_u}(0)\sum_{i=1}^4
\frac{G^{\phi\chi}_{k_i}(0)}{G^{\chi\chi}_{k_i}}\nn\\
&&\hspace{-10pt}+4
\lambda_3^2G^{\chi\chi}_{k_u}\sum_{i=u_1,u_2}\sum_{j=u_3,u_4}\frac{G^{\phi\chi}_{k_i}(0)}{G^{\chi\chi}_{k_i}}
\frac{G^{\phi\chi}_{k_j}(0)}{G^{\chi\chi}_{k_j}}
+4\lambda_3^2
\frac{G^{\phi\chi}_{k_u}(0)G^{\phi\chi}_{k_u}(0)}{G^{\chi\chi}_{k_u}}\Bigg)G^{\chi\chi}_{k_u}
\(\prod_{i=1}^4 G^{\chi \chi}_{k_i}\)\,.\nn
\ea
As shown symbolically in Fig.~\ref{4pt}, after appropriate recombination of these different contributions, we
can reexpress this four-point function as a finite product of these two
renormalized three-point function plus a divergent piece which
fixes the renormalization of the coupling $\beta_4$:
\ba
\hspace{-20pt}G^{\chi\chi\chi\chi}_{k_1,k_2,k_3,k_4}
&=&\sum_{\rm{perm}_u}\Bigg\{\frac{1}{G^{\chi\chi}_{k_u}}\mathcal{A}_{u}
+4(-i)^2\lambda_3^2\left[G^{\phi\phi}_{k_u}(0,0)-\frac{G^{\phi\chi}_{k_u}(0)G^{\phi\chi}_{k_u}(0)}{G^{\chi\chi}_{k_u}}
\right]\(\prod_{i=1}^4 G^{\chi \chi}_{k_i}\)\Bigg\}\nn\\&&+4!(-i)\beta_4\(\prod_{i=1}^4 G^{\chi
\chi}_{k_i}\)\nn\\
\label{4ptfunc}
&=&\sum_{\rm{perm}_u}\Bigg\{\frac{1}{G^{\chi\chi}_{k_u}}\mathcal{A}_{u}
+\left[4(-i)^2\lambda_3^2 \tilde D_{k_u}(0,0)+(-i)\frac{4!}{3}\beta_4\right]\(\prod_{i=1}^4 G^{\chi
\chi}_{k_i}\)\Bigg\}.
\ea
All terms in the previous expressions are finite apart from the ones
in square brackets. The divergence of this term can once again be
absorbed into $\beta_4$ using the following appropriate
renormalization
\ba
\label{beta4}
\beta_4(\Lambda)-\frac{i}2\lambda_3^2(\Lambda)\tilde D_k(0,0)&=&
\beta_4(\Lambda)-\frac{\lambda_3^2(\Lambda)}{4\pi\alpha}\frac{\log \frac{\Lambda}{k}}{1+\frac{\lambda_2(\Lambda)}{2\pi \alpha}\log\frac\Lambda
k}\nn\\
&=&\beta_4(\mu)-\frac{\lambda_3^2(\mu)}{4\pi\alpha}\frac{\log
\frac{\mu}{k}}{1+\frac{\lambda_2(\mu)}{2\pi
\alpha}\log\frac\mu
k}\,,
\ea
\ie the renormalized coupling $\beta_4$ must flow as
\ba
\mu \partial_\mu \beta_4(\mu)=\frac{\lambda_3^2(\mu)}{4 \pi
\alpha}.
\ea
In other words, as soon as cubic interactions between the
bulk and the brane field are introduced, a quartic interaction for
the brane field is spontaneously generated classically. This is
familiar for standard EFT.

We can also carefully check using exactly the same technique as previously
that this renormalization of the coupling $\beta_4$  also ensures
that all remaining four-point functions $\langle\phi
\chi^3\rangle$, $\langle\phi^2 \chi^2\rangle$, $\langle\phi^3 \chi
\rangle$ and  $\langle\phi^4\rangle$ are completely finite
classically (provided the bulk field is evaluated away from the
brane),
\ba
G^{\phi\chi\chi\chi}_{k_1,k_2,k_3,k_4}(r)
&=&
\sum_{\rm{perm}_u}\Bigg\{\frac{1}{G^{\chi\chi}_{k_u}}
G^{\phi\chi\chi}_{k_{u_1},k_{u_2},k_u}(r)G^{\chi\chi\chi}_{k_{u_3},k_{u_4},k_u}\nn\\
&&-4\left[\lambda_3^2 \tilde
D_{k_u}(0,0)+2i\beta_4\right]G^{\phi\chi}_{k_1}(r)\(\prod_{i=2}^4
G^{\chi
\chi}_{k_i}\)\Bigg\}\nn\\
G^{\phi\phi\chi\chi}_{k_1,k_2,k_3,k_4}(r_1,r_2)
&=&\frac{1}{G^{\chi\chi}_{k_1+k_2}}
G^{\phi\phi\chi}_{k_1,k_2,k_1+k_2}(r_1,r_2)G^{\chi\chi\chi}_{k_3,k_4,k_1+k_2}\nn\\
&&+\frac{1}{G^{\chi\chi}_{k_1+k_3}}G^{\phi\chi\chi}_{k_1,k_3,k_1+k_3}(r_1)G^{\phi\chi\chi}_{k_2,k_4,k_1+k_3}(r_2)
+(3\leftrightarrow4)\nn\\
&&-\Bigg(4\left[\lambda_3^2 \tilde
D_{k_1+k_3}(0,0)+2i\beta_4\right]+(3\leftrightarrow4)\nn\\
&&+4\left[\lambda_3^2 \tilde
D_{k_1+k_2}(0,0)+2i\beta_4\right]\Bigg)
\(\prod_{i=1}^2 G^{\phi \chi}_{k_i}(r_i)\)\(\prod_{i=3}^4 G^{\chi
\chi}_{k_i}\)\nn
\ea
\ba
G^{\phi\phi\phi\chi}_{k_1,k_2,k_3,k_4}(r_1,r_2,r_3)
&=&
\sum_{\rm{perm}_u}\Bigg\{\frac{1}{G^{\chi\chi}_{k_u}}
G^{\phi\phi\chi}_{k_{u_1},k_{u_2},k_u}(r_{u_1},r_{u_2})G^{\phi\chi\chi}_{k_{u_3},k_4,k_u}(r_{u_3})\nn\\
&&-4\left[\lambda_3^2 \tilde
D_{k_u}(0,0)+2i\beta_4\right]\(\prod_{i=1}^3
G^{\phi\chi}_{k_i}(r_i)\) G^{\chi
\chi}_{k_4}\Bigg\}\nn\\
G^{\phi\phi\phi\phi}_{k_1,k_2,k_3,k_4}(r_1,r_2,r_3,r_4)
&=&
\sum_{\rm{perm}_u}\Bigg\{\frac{1}{G^{\chi\chi}_{k_u}}
G^{\phi\phi\chi}_{k_{u_1},k_{u_2},k_u}(r_{u_1},r_{u_2})G^{\phi\phi\chi}_{k_{u_3},k_{u_4},k_u}(r_{u_3},r_{u_4})\nn\\
&&-4\left[\lambda_3^2 \tilde
D_{k_u}(0,0)+2i\beta_4\right]\(\prod_{i=1}^4
G^{\phi\chi}_{k_i}(r_i)\) \Bigg\}\nn\,.
\ea
Each of these four-point functions introduces a combination of the
couplings which is completely finite once they have been
renormalized as specified previously
(\ie $\lambda_3^2+\tilde D_k(0,0)+2i\beta_4$ is finite).
This non-trivial check ensures
that our proposal makes sense at least at the classical level up to the
four-point function. Before discussion the general renormalizability
of this theory for higher point functions, we present in what follows
an insight into the situation at the quantum level, \ie when loops are
taken into account.

\subsection{Loops}
At the loop level, we expect from standard field theory in four dimensions, that UV divergences will arise from
the momentum integral over the loop. However no further divergences
arise from the codimension-two nature of the theory, and the
counterterms required to absorb the divergences are thus the usual
one of four-dimensional field theory.

To start with, we concentrate on the two-point function of the brane field $\chi$.
At first order in loops, (second order for $\beta_3$ and
$\lambda_3$) one has
\ba
\langle \chi \chi \rangle_{1\underline{\ }\,\rm{loop}} =
\int \frac{\d^4 p}{(2\pi)^4}\ \mathcal{I}_{\rm{loop}}(p,k)\ G^{\chi \chi}_k G^{\chi
\chi}_{k}\,,
\ea
with the integrand $\mathcal{I}_{\rm{loop}}(p,k)$ being the sum over
the different loop configurations:
\ba
\hspace{-20pt}\mathcal{I}_{\rm{loop}}(p,k)\hspace{-5pt}&=&\hspace{-5pt}
\frac{(-i)^2}{2!}G^{\chi \chi}_pG^{\chi \chi}_{k-p}\(
6 \beta_3
-2 i \lambda_3\lambda\(\tilde D_k +\tilde D_p +\tilde
D_{k-p}\)\)^2
\\
&&\hspace{-5pt}
\frac{(-i)^2}{2!}G^{\chi \chi}_kG^{\chi \chi}_0\(
6 \beta_3
-2 i \lambda_3\lambda\(\tilde D_k +\tilde D_p +\tilde
D_0\)\)^2\nn \\
&&\hspace{-5pt}+4 \lambda_3^2\(G^{\chi\chi}_p
\tilde D_{k-p}+G^{\chi\chi}_{k-p}\tilde D_{p}+G^{\chi\chi}_p \tilde D_{0}\)
+ (-i)\frac{4!}{2}\beta_4 G^{\chi\chi}_p\,, \nn
\ea
where for simplicity we have used the notation $\tilde D_k\equiv \tilde
D_k(0,0)$. This expression can be most easily interpreted as
finite products of three-point functions and extra terms as
symbolized in Fig. \ref{FigLoops}
\ba
\label{loop1}
\mathcal{I}_{\rm{loop}}(p,k) &=&
-\frac 12 \frac{(G^{\chi\chi\chi}_{k,p,k-p})^2}{G^{\chi\chi}_{p}G^{\chi\chi}_{k-p}(G^{\chi\chi}_{k})^2}
-\frac 12 \frac{G^{\chi\chi\chi}_{k,k,0}G^{\chi\chi\chi}_{p,p,0}}{G^{\chi\chi}_{p}G^{\chi\chi}_{0}(G^{\chi\chi}_{k})^2}\\
&&-2
\left[\lambda_3^2 \tilde
D_{p}+2i\beta_4\right]G^{\chi\chi}_{k-p}
-2\left[\lambda_3^2 \tilde
D_{k-p}+2i\beta_4\right]G^{\chi\chi}_{p}\nn\\
&&-2
\left[\lambda_3^2 \tilde
D_{0}+2i\beta_4\right]G^{\chi\chi}_{p}
\nn\,.
\ea

\begin{figure}[h]
\begin{picture}(150,100)(-50,-10)\thicklines
%\large
\crr
\put (-40,60){\makebox(0,0)[l]{$\langle \chi^2 \rangle_{1\underline{\ }\,\rm{loop}}
\   \color{black}= - i \, \frac{4!}{2} \, \crr \beta_4$}}
\color{black}
\put (125,60){\makebox(0,0)[l]{$+\ \ \frac{(-i)^2}{2} \ \Bigg[$}}
\put (248,60){\makebox(0,0)[l]{$+\ 2. 4\, \cb \lambda_3^2$}}
\put (20,10){\makebox(0,0)[l]{$+\ \ \frac{(-i)^2}{2} \ \Bigg[$}}
\put (170,10){\makebox(0,0)[l]{$+\  4\, \cb \lambda_3^2$}}
\color{black}
\put (365,60){\makebox(0,0)[l]{$\Bigg]$}}
\put (300,10){\makebox(0,0)[l]{$\Bigg]$}}
\crr
\put(65,60) {\line(1,0){52}}
\put(91,68){\circle{16}}
\put(78,60){\circle*{4}}
\put(104,60){\circle*{4}}
\put(91,76){\circle*{4}}
\put (180,60){\line(1,0){60}}
\put (210,60){\oval(26,20)[t]}
\put (186,60){\circle*{4}}
\put (234,60){\circle*{4}}
\put (210,70){\circle*{4}}
\put (210,60){\circle*{4}}
\color{white}
\put (196,60){\circle*{8}}
\put (224,60){\circle*{8}}
\color{blue}
\put (192,56){\line(0,1){8}}
\put (200,56){\line(0,1){8}}
\put (192,56){\line(1,0){8}}
\put (192,64){\line(1,0){8}}
\put (220,56){\line(0,1){8}}
\put (228,56){\line(0,1){8}}
\put (220,56){\line(1,0){8}}
\put (220,64){\line(1,0){8}}
\multiput(315,60)(2,0){13}{\circle*{1.45}}\crr
\put (295,60){\line(1,0){20}}
\put (341,60){\line(1,0){20}}
\put (328,60){\oval(26,20)[t]}
\put (305,60){\circle*{4}}
\put (328,70){\circle*{4}}
\put (351,60){\circle*{4}}
%%
%%%%%%%%%%%
%%
%\cb \multiput(50,10)(2,0){8}{\circle*{1.45}}\crr
\put(90,0) {\line(1,0){60}}
\put(120,0){\line(0,1){22}}
\put(120,30){\circle{16}}
\put(100,0){\circle*{4}}
\put(140,0){\circle*{4}}
\put(120,38){\circle*{4}}
\put(120,11){\circle*{4}}
\cb \put (116,-4){\line(0,1){8}} \put (124,-4){\line(0,1){8}} \put (116,-4){\line(1,0){8}} \put (116,4){\line(1,0){8}}
\put (116,18){\line(0,1){8}} \put (124,18){\line(0,1){8}} \put (116,18){\line(1,0){8}} \put (116,26){\line(1,0){8}}
\color{white} \put (120,0){\circle*{7}} \put (120,22){\circle*{7}}\crr
\put(220,0) {\line(1,0){60}}
%\put(250,0){\line(0,1){22}}
\put(250,30){\circle{16}}
\put(230,0){\circle*{4}}
\put(270,0){\circle*{4}}
\put(250,38){\circle*{4}}
\cb \multiput(250,0)(0,2){12}{\circle*{1.45}}\crr
\end{picture}
\caption{One-loop corrections to the two-point function. On both lines, the second diagram is finite, while the first and third diagrams
contain logarithmic divergences that will not cancel each other.}
\label{FigLoops}
\end{figure}

To clarify the discussion, we denote by $\Lambda$, the cutoff scale
associated with the codimension-two brane thickness, or equivalently
with the integration over the momentum along the extra dimensions,
while $\Delta$ designates the standard four-dimensional momentum cut-off scale,
\ie in \eqref{loop1} the loop integration is cutoff at the scale
$\Delta$: $\int \d^4 p \sim \int_0^\Delta \d p p^3$. Clearly these
two scales could be associated with one another, simply representing
the scale at which UV physics becomes important. However for sake of
simplicity, we distinguish for now between these two quantities and assume
that in general they could be different. In this scenario, the couplings are then
flowing along two distinct directions $\Lambda$ and $\Delta$, and we
focus our attention on the flow along the $\Lambda$ direction.

On simple dimension grounds, we expect that the loop
integral over $p$ will diverge logarithmically, and the one-loop
contribution to the two-point function has thus a cutoff dependence
of the form $\log \Delta$ which should be absorbed by introduction
of a mass counterterm of the form $\delta m^2 \sim \log \Delta$.
This is standard procedure in four-dimensional field theory.

We are however more concern here on the dependence of the other cutoff $\Lambda$. $\lambda_3$ and $\beta_3$ have
been renormalized in \eqref{beta3_2},
such that the three-point function $G^{\chi\chi\chi}$ is finite so
the first line of \eqref{loop1} is clearly finite. Although the rest
of expression \eqref{loop1} includes divergent terms of the form $\tilde
D_k(0,0)$, the combination involved $\left[\lambda_3^2 \tilde
D_{k}+2i\beta_4\right]$ is precisely the combination that
appeared in the expression of four-point function (\ref{4ptfunc},
\ref{beta4}), and is thus also finite.

Notice furthermore that the one-loop correction to the two
remaining two-point functions $\langle \phi(r) \chi \rangle_{1\underline{\ }\,\rm{loop}}$ and
 $\langle \phi(r) \phi(r') \rangle_{1\underline{\ }\,\rm{loop}}$ will be
 also be finite in the thin-brane limit (once the loop divergences associated with $\Delta$ have been
 taken care of), as shown explicitly for the four-point function.

However a non-trivial feature emerges from the computation of the one-loop
corrections. The expression \eqref{loop1} involves terms of
the form $\tilde D_0(0,0)$ which also diverges logarithmically, but
this time this divergence is instead associated with a IR behaviour.
In this toy-model, the origin of this IR divergence is related to
the fact that the bulk field $\phi$ is massless, but would disappear
as soon as a small mass $m_\phi^2$ was introduced. However, if the
bulk field is to mimic the graviton, this field should remain massless. Physically, such IR divergences
can be removed in the same way as in quantum
electrodynamics, see ref.~\cite{Weinberg:1965nx}.

\subsection{Is the theory renormalizable?}

To complete this section, we argue that this scalar field toy-model is
renormalizable, provided that only relevant and marginal operators are
considered. Since no coupling of the form $\beta_N \chi^N$, $(N\ge
5)$ is introduced, any further $N$-point function will necessarily be
composed of only reducible diagrams and will thus be expressible in
terms of lower-dimensional $n$-point functions ($n\le4$.) Since we
have shown that all of these $n$-point functions are finite at the
classical level, any further $N$-point function will thus
automatically be finite, without any further counterterms.

We make this argument more concrete by exploring the five-point
function, and showing explicitly that it remains finite in the
thin-brane limit if $\beta_4$ is renormalized as in
\eqref{beta4}. A completely general argument for an arbitrary
$N$-point function can be found in appendix~\ref{appendix generalcase}.
In particular we show that the same will remains true at
the level of any $N$-point function.
The theory will thus be renormalizable, as one can expect from standard
four-dimensional field theory intuition.

The contributions to the five-point function are symbolized in
Fig.\ref{Fig5pt}.
\begin{figure}[h]
\begin{picture}(150,70)(-20,40)\thicklines
%\large
\crr
\put (-20,60){\makebox(0,0)[l]{$\langle \chi^5 \rangle
\  \, \color{black}= \, \frac{(- i)^2}{2!} \  4.5!\  \crr \beta_4$}}
\tiny
\color{black}
%\put (28,50){\makebox(0,0)[l]{10 perms.}}
%\put (160,50){\makebox(0,0)[l]{15 perms.}}
\normalsize
\put (130,60){\makebox(0,0)[l]{$+ \ \frac{(-i)^3}{3!}\,  \frac{3!  5!}{2^3} \ \Bigg[$}}
\put (270,60){\makebox(0,0)[l]{$+\ 2^3\, \cb \lambda_3^2$}}
\put (360,60){\makebox(0,0)[l]{$\Bigg]$}}
%\put (125,10){\makebox(0,0)[l]{$+\ \frac{(-i)^2}{2}\frac 12 \ \Bigg[$}}
%\put (248,10){\makebox(0,0)[l]{$+\ 2. 4\, \cb \lambda_3^2 \lambda$}}
%\color{black}
%\put (365,60){\makebox(0,0)[l]{$\Bigg]$}}
%\put (375,10){\makebox(0,0)[l]{$\Bigg]$}}
\crr
\put(90,64) {\line(-1,1){10}}
\put(90,56) {\line(-1,-1){10}}
\put(94,60) {\line(1,0){30}}
\put(114,60){\line(1,1){10}}
\put(114,60){\line(1,-1){10}}
\put(85,69){\circle*{4}}
\put(85,51){\circle*{4}}
\put(104,60){\circle*{4}}
\put(120,60){\circle*{4}}
\put(120,66){\circle*{4}}
\put(120,54){\circle*{4}}
\color{blue}
\put (86,56){\line(0,1){8}}
\put (94,56){\line(0,1){8}}
\put (86,56){\line(1,0){8}}
\put (86,64){\line(1,0){8}}
\crr
\put(210,64) {\line(-1,1){10}}
\put(210,56) {\line(-1,-1){10}}
\put(234,64) {\line(0,1){10}}
\put(258,64) {\line(1,1){10}}
\put(258,56) {\line(1,-1){10}}
\put(238,60) {\line(1,0){16}}
\put(214,60) {\line(1,0){16}}
\put(205,69){\circle*{4}}
\put(205,51){\circle*{4}}
\put(222,60){\circle*{4}}
\put(234,69){\circle*{4}}
\put(246,60){\circle*{4}}
\put(263,69){\circle*{4}}
\put(263,51){\circle*{4}}
\color{blue}
\put (206,56){\line(0,1){8}}
\put (214,56){\line(0,1){8}}
\put (206,56){\line(1,0){8}}
\put (206,64){\line(1,0){8}}
\put (230,56){\line(0,1){8}}
\put (238,56){\line(0,1){8}}
\put (230,56){\line(1,0){8}}
\put (230,64){\line(1,0){8}}
\put (254,56){\line(0,1){8}}
\put (262,56){\line(0,1){8}}
\put (254,56){\line(1,0){8}}
\put (254,64){\line(1,0){8}}
\crr
\put(320,64) {\line(-1,1){10}}
\put(320,56) {\line(-1,-1){10}}
\put(324,60) {\line(1,0){15}}
\put(349,60){\line(1,1){10}}
\put(349,60){\line(1,-1){10}}
\put(339,60){\line(0,1){10}}
\put(315,69){\circle*{4}}
\put(315,51){\circle*{4}}
%\put(334,60){\circle*{4}}
\put(331,60){\circle*{4}}
\put(355,66){\circle*{4}}
\put(355,54){\circle*{4}}
\put(339,65){\circle*{4}}
\color{blue}
\put (316,56){\line(0,1){8}}
\put (324,56){\line(0,1){8}}
\put (316,56){\line(1,0){8}}
\put (316,64){\line(1,0){8}}
\cb \multiput(339,60)(2,0){5}{\circle*{1.45}}\crr
\crr
%
%%%%%%%%%%%
%%
\end{picture}
\caption{Classical contributions to the five-point function. The second diagram is finite, while the first and third diagrams
contains logarithmic divergences that compensate each other. This five-point function is therefore finite.}
\label{Fig5pt}
\end{figure}

Summing these diagrams, we obtain
\ba
G^{\chi^5}_{k_1,\cdots,k_5}&=&
\frac{(-i)^3}{3!}\, \frac{ 3! 5!}{2^3}\, \frac{G^{\chi\chi\chi}_{k_1, k_2, (k_1+k_2)}G^{\chi\chi\chi}_{k_3, (k_1+k_2),(k_4+k_5)}
G^{\chi\chi\chi}_{(k_4+k_5), k_4,
k_5}}{G^{\chi\chi}_{k_1+k_2}G^{\chi\chi}_{k_4+k_5}}\\
&&
+\(\frac{(-i)^3}{3!}\, \frac{ 3! 5!}{2^3} \, 2^3
\lambda_3^2\tilde
D_{k_1+k_2}
+\frac{(-i)^2}{2!}\,4. 5!\,
\beta_4\)G^{\chi\chi\chi}_{(k_4+k_5),k_4,k_5}\prod_{i=1}^3G^{\chi\chi}_{k_i}
\nn\,\\
&&\hspace{-70pt}=\ 5! i\Bigg[
\frac{G^{\chi\chi\chi}_{k_1, k_2,
(k_1+k_2)}G^{\chi\chi\chi}_{k_3,
(k_1+k_2),(k_4+k_5)}}{G^{\chi\chi}_{k_1+k_2}G^{\chi\chi}_{k_4+k_5}}
+\(\lambda_3^2\tilde
D_{k_1+k_2}
+2i \beta_4\)\prod_{i=1}^3G^{\chi\chi}_{k_i}\Bigg]G^{\chi\chi\chi}_{(k_4+k_5), k_4,
k_5}\nn\,,
\ea
where for simplicity we have used a specific momentum configuration, but the counting takes in account all possible
permutations.

The first line of the pervious expression is trivially finite, while
the second line is finite only if the terms proportional to
$\beta_4$ and $\lambda_3^2 \tilde D_k(0,0)$ contribute with
appropriate coefficients. As can be seen in the third line of this
expression, the contribution from these terms is also finite as
$\beta_4$ is renormalized precisely so as to have $\lambda_3^2\tilde
D_{k}+2i \beta_4$ finite. The five-point function is therefore
finite at the classical level and no counterterms ought to be added.
This result will remain valid for any other five-point functions (\ie including
the ones with bulk external fields $\langle \phi(r)
\chi\chi\chi\chi\rangle$, etc.), as well as for any higher $N$-point
function and their loop corrections, (see appendix~\ref{appendix generalcase}).
This represents a highly non-trivial check and leads to the
conclusion that the theory is completely renormalizable against divergences associated with
the codimension-two source.

%\part{Physical implications}

%%%%%%%%%%%%%%%%%%%%%%%%%%%%%%%%%%%%%%%%%%%%%%%
%%%%%%%%%     COUPLING WITH EM

\section{Electromagnetism on a codimension-two brane}
\label{section em}
% \subsection{Detection of gravitational waves}

In this section, we consider the more physical scenario of
a massless gauge field $A_\mu$ confined to the brane and coupled to
gravity. This represents a more realistic framework to study the
coupling between gravity and electromagnetism.
We start with the following six-dimensional action
\ba
S^{(\rm{em})}=\int \d^6 x \sqrt{-g}\left[\frac{1}{2\kappa^2}R^{(6)}-
\delta^{2}(y)\ \frac 14 F_{\mu\nu}F^{\mu\nu}\right]\,,
\ea
with $F_{\mu\nu}=\partial_\mu A_\nu-\partial_\nu A_\mu$. In what
follows we work at linear order in perturbations around a flat
conical background:
\ba
\d s^2= g_{a b} \d x^a \d x^b=\d r^2+r^2 \d
\theta^2+\(\eta_{\mu\nu}+h_{\mu\nu}\)\d x^\mu \d x^\nu\,,
\ea
where we work in de Donder gauge, $h^{\mu}_{\nu\, , \mu}=\frac 12 h^\mu_{\mu\,
,\nu}$. Since the stress-energy for radiation is transverse, we will
have $h^\mu_{\ \mu}=0$. The Einstein's equations impose
\ba
G_{\mu \nu}&=& \kappa^2 T_{\mu\nu}^{\rm{em}}\nn\\
-\frac 12 \Box^{(6)} h_{\mu\nu}&=& -\kappa^2\frac{\delta(r)}{2r \pi \alpha}\(F_{\mu\alpha}F_{\nu}^{\ \alpha}-\frac 14 F^2 \eta_{\mu
\nu}\)\,.
\ea
Using results from the previous sections, we know that $\tilde
h_{\mu\nu}$ will diverge logarithmically when evaluated at $r=0$
\ba
\label{induced h}
h_{\mu \nu}(0)= \frac{\kappa^2}{4\pi \alpha} \(\Gamma+\log \frac{k \epsilon}{2}\)(F_{\mu\alpha}F_\nu^{\ \alpha}-\frac 1 4 F^2 \eta_{\mu
\nu})\,,
\ea
where $\Gamma$ is the Euler number and $\epsilon \rightarrow 0$ represents the thin-brane limit.
This will affect the equation of motion for the photon:
\ba
\nabla_\mu F^{\mu \nu}&=&\(\eta ^{\mu \tilde \mu}-h^{\mu \tilde \mu}\)\(\eta^{\nu \tilde\nu}-h^{\nu
\tilde\nu}\)\(\partial_\mu F_{\tilde \mu \tilde \nu}-\Gamma^\alpha _{\mu \tilde \mu}F_{\alpha \tilde \nu}-\Gamma^\alpha _{\mu \tilde \nu}F_{ \tilde \mu \alpha}\)\nn\\
&=&
\label{Maxwell1}
\partial_\mu F^{\mu \nu}-\partial_\mu \(h^{\alpha \nu}F^\mu_{\ \ \alpha}-h^{\alpha \mu}F
^\nu_{\ \ \alpha}\)=0\,,
\ea
where in the second line, all index raising is performed with
respect to the background flat metric $\eta^{\alpha \beta}$. We
remember that in the previous expression, $h_{\alpha \beta}$
represents the induced value of the metric perturbation evaluated on
the brane and thus diverges logarithmically in the thin-brane limit.
If this was the end of the story, then photons would be very
sensitive to the brane thickness $\epsilon$ even at low-energy.
However, we have learned from section \ref{Section free theory}, that
as soon as a coupling $\lambda$ is introduced between brane and bulk
fields, this spontaneously generates a mass term $m^2$ for the brane
field at the classical level. The situation is no different here,
and the logarithmic divergence of $h$ on the brane will
spontaneously generate $F^4$ terms on the brane. More
precisely, let us consider the Euler-Heisenberg brane action
\ba
S_{(\rm{brane})}=-\int \d x^4 \sqrt{-q} \left[ \frac 14
F_{\mu\nu}F^{\mu\nu}+\frac{\gamma_1}{8} \(F_{\mu\nu}F^{\mu\nu}\)^2
+\frac{\gamma_2}{8} F_{\mu \nu}F^\mu_{\ \ \alpha}F_\beta^{\ \ \alpha}F^{\beta
\nu}\right]\,,
\ea
so that the Maxwell's equations \eqref{Maxwell1} are modified to
\ba
\partial_\mu F^{\mu  \nu}-\partial_\mu \(h^{\alpha \nu}F^\mu_{\ \ \alpha}-h^{\alpha \mu}F
^\nu_{\ \ \alpha}\)-\gamma_1 \partial_\mu\(F^2 F^{\mu
\nu}\)-\gamma_2 \partial_\mu \(F^\mu_{\ \ \alpha}F^\alpha_{\ \
\beta}F^{\nu\beta}\)=0.
\ea
Substituting the expression \eqref{induced h} for the perturbed metric on the
brane, we obtain
\ba
\partial_\mu F^{\mu \nu}
&+&\(\frac{\kappa^2}{8\pi \alpha}\(\Gamma+\log\frac{k
\epsilon}{2}\)-\gamma_1\)
\partial_\mu \( F^2 F^{\mu \nu}\)\nn\\
&-&\(\frac{\kappa^2}{2\pi \alpha}\(\Gamma+\log\frac{k \epsilon}{2}\)+\gamma_2\)
\partial_\mu \(F^\mu_{\ \ \alpha}
F^\alpha_{\ \ \beta} F^{\nu \beta}\)=0\,.
\ea
The divergence of the graviton can thus be absorbed in the two couplings $\gamma_1$ and
$\gamma_2$:
\ba
\gamma_1(\mu)=\gamma_1(\epsilon)+\frac{\kappa^2}{8 \pi \alpha} \log
\frac{\mu}{\epsilon}\hspace{20pt}\rm{and}\hspace{20pt}
\gamma_2(\mu)=\gamma_2(\epsilon)-\frac{\kappa^2}{2 \pi \alpha} \log
\frac{\mu}{\epsilon}\,,
\ea
leading to the following RG flows
\ba
\mu \partial_\mu \gamma_1(\mu)
=-\frac 1 4\ \mu \partial_\mu \gamma_2(\mu)
=\frac{\kappa^2}{8\pi \alpha}\,.
\ea
The generation of $F^4$ terms on the brane at the classical level
ensures that a photon confined to a codimension-two brane and interacting with a gravitational
wave will not evolve in a regularization-dependent way. In
particular, this mechanisms ensures that at low-energy, there is a
well defined thin-brane limit description of the codimension-two
brane. Of course once, $F^4$ terms are introduced, they will in turn introduce divergences on the
brane, which should be absorbed with higher order terms. The proper
finite theory will hence include a infinite series.

%%%%%%%%%%%%%%%%%%%%%%%%%%%%%%%%%%%%%%%%%%%%%%%
%%%%%%%%%     KINETIC TERM ON THE BRANE

\section{Localized Kinetic Terms}
\label{section kinetic terms}
As a last intriguing physical implication, we consider in this
section the consequences for localized kinetic terms on the brane.
Localized kinetic terms are of importance when considering
brane-induced Einstein-Hilbert terms, where the action is typically
of the form
\ba
S=\frac{M_{(d)}^{d-2}}{2} \int \d ^{d} x \sqrt{-g_d}\, R_{(d)}
+\int \d^4 x\sqrt{-g_4}
\,\(\frac{M_{(4)}^{2}}{2}R_{(4)}+\mathcal{L}_{\rm{matter}}\).
\ea
The induced Einstein-Hilbert term $R_{(4)}$ term is expected to be
spontaneously generated at the quantum level, and represents a
natural mechanism to localize gravity on a four-dimensional surface
when the extra dimensions are flat and infinite
\cite{Dvali:2000rv,deRham:2006hs}. Such models also represent a physical
realization of the degravitation process (see ref.~\cite{Dvali:2007kt}), since in
such scenarios gravity becomes fully higher-dimensional at long
wavelengths, hence providing a potential explanation for
the observed value of the cosmological constant. Such models are
also enriched with an additional interesting feature namely the
possibility of having self-accelerating branches
\cite{Deffayet:2001pu} (see refs.~\cite{deRham:2006pe} for
ghost-free realizations).

Although such models are usually considered in the context of one
large extra dimension, a simultaneous resolution of the Hierarchy
problem usually requires at least two extra dimensions
\cite{Arkani-Hamed:1998rs}, and the degravitation observed in the
presence of only one extra dimension is only marginal. Understanding
this scenario in the presence of two extra dimensions is therefore
an important next step. In what follows, we examine the consequences
of such kinetic terms in a scalar field toy-model.

We consider a massless scalar field $\phi$ living in a six-dimensional
flat space-time with induced kinetic terms on a codimension-two
brane
\ba
S=-\int \d^6x\left[\frac12 (\partial_a \phi)^2+ \delta^2(y) \, \phi
\, f(\Box) \, \phi \right]\,,
\ea
where $\Box$ represents the four-dimensional d'Alembertian $\Box=\partial^\mu
\partial_\mu$. In order to recover on the brane the standard Klein-Gordon
equation for the scalar field in the infrared, $\(\Box
+m^2\)\phi=0$, we require that only positive powers of $\Box$ be
present in $f$. In particular we write
\ba
f(\Box)= \sum_{n\ge 0} c_n
\(\ell^2 \Box \)^n,
\ea
where $\ell$ is an arbitrary length scale (we recall that in this
six-dimensional formalism, $\phi$ has dimension mass squared and thus $f$ ought
to be dimensionless). In particular, $c_0$ represents the
dimensionless coupling $\lambda_2$ that was considered in section \ref{Section free
theory}.
From that section, we know that the scalar field will be well-defined away from the brane only
if the induced couplings on the brane (in this case the function $f$) are renormalized and flow as
\ba
\mu \partial_\mu f(\Box)=\frac 1{2\pi\alpha} f^2(\Box)\,.
\ea
In terms of the coefficients $c_n$, this implies
\ba
\label{cn'}
\mu \partial_\mu c_n(\mu)=\frac 1{2\pi\alpha} \sum_{u=0}^n c_{n-u}(\mu)c_u(\mu)\,.
\ea
This has important consequences for these kind of theories, in
particular, we are not free to choose a brane induced function
of the form $f(\Box)=(m^2+\Box)$, as higher curvature terms will
spontaneously be generated at the tree-level. As soon as a mass term
$c_0$ and kinetic term $c_1$ are present, all the other couplings
$c_n$ will flow in a non-trivial way. The solution for the two first
terms is of the form
\ba
c_0(\mu)&=&\frac{\bar c_0}{1-\frac{1}{2\pi\alpha}\bar c_0 \log
\mu}\\
c_1(\mu)&=&\frac{\bar c_1}{\(2\pi \alpha-\bar c_0 \log
\mu\)^2}=\beta c_0(\mu)^2\,,
\ea
where $\beta=\bar c_1 / 2\pi \alpha \bar c_0^2$ is a dimensionless
parameter.

As an example, one can choose
a particular solution of \eqref{cn'}, for which
\ba
c_n(\mu)=\beta^n c_0(\mu)^{n+1}\,,
\ea
and so
\ba
f(\Box)=\frac{c_0(\mu)}{1-\beta \, c_0(\mu) \Box}\,.
\ea
Notice that in this formalism, the function of the kinetic term is
fixed by the renormalization conditions, and very few parameters can
actually be tuned arbitrarily. In this sense this represents a much
more satisfying candidate for theories of modified gravity than for
instance $f(R)$ gravities, (see ref.~\cite{Copeland:2006wr} for a review on such theories). As a natural extension, one should
understand the cosmology for such a scenario and possibly the
different signatures which could allow for the discrimination of
codimension-two models.

%%%%%%%%%%%%%%%%%%%%%%%%%%%%%%%%%%%%%%%%%%%%%%%
%%%%%%%%%     CONCLUSION

\section{Conclusions}
\label{section conclusion}

We have analyzed the coupling between bulk fields living in a
six-dimensional flat space-time and brane fields confined onto a
four-dimensional surface. Due to these couplings, logarithmic divergences that arise when
evaluating the bulk field on the brane generically propagate into
the brane field. In this paper, we have presented a consistent
renormalization mechanism at tree and one-loop level that
removes any divergences simultaneously in the brane field and bulk
field when the latter is evaluated away from the brane.
We have also shown that any five-point function is finite at the
classical level without the addition of any further counterterm, and
demonstrated that this remains true for any $N$-point function at any
order in the loop expansion,
thus proving the  renormalizability of the theory.
We also point out the presence of IR divergences in the loop
diagrams which can be dealt with the same way as for quantum
electrodynamics.

The same principle can be applied to more complex theories such as
electromagnetism in curved space-time, for which the same
prescription remains completely valid. In particular we show that at
tree-level, the coupling of electromagnetism to gravity generates a
quartic term for the form field in the action, giving rise to a
Euler-Heisenberg Lagrangian which is usually only
generated via quantum corrections.

As another physical application, we have also explored the
consequences for localized kinetic terms which are relevant in
scenarios such as the DGP model. In particular we show that as soon as
a kinetic term are induced on the brane, one cannot prevent
for the generation of an infinite series of higher order terms. This
provides a natural modification of gravity on the brane which might
have potential interesting signatures.

To our knowledge, this prescription is the only one to date capable of
making sense of sources on codimension-two branes in a
regularization-independent way and providing a way to derive the
low-energy effective theory on such objects in a regime where
interactions with the bulk cannot be ignored.

Since the main objective of this paper was the establishment of a
consistent framework to study sources on codimension-two branes,
implications for braneworld physics have only been
superficially addressed. Armed with these new tools,
extensions to more physical scenarios will however be of great
importance. Understanding the relevance of our results for
electromagnetism and theories with induced gravity terms were beyond
the scope of this paper but will present interesting extensions.
More realistic interactions from the standard model
would also be intriguing, and in particular
consequences to the Higgs physics in six-dimensional scenarios, such
as the SLED, \cite{Beauchemin:2004zi} should be understood in more
detail.

\section*{Acknowledgements}

The author wishes to thank Rouzbeh Allahverdi, Cliff Burgess, Stefan Hofmann, Justin Khoury,  Federico Piazza, Andrew Tolley, Mark Wise and Mark
Wyman for their contribution throughout the development of this
work.
Research at McMaster is supported by the Natural
Sciences and Engineering Research Council of Canada.
Research at Perimeter Institute for Theoretical Physics is supported
in part by the Government of Canada through NSERC and by the
Province of Ontario through MRI.

%%%%%%%%%%%%%%%%%%%%%%%%%%%%%%%%%%%%%%%%%%%%%%%
%%%%%%%%%     APPENDIXES

\appendix

\renewcommand{\thesection}{}

\section{\hspace{-20pt}Appendix}

\renewcommand{\thesection}{\Alph{section}}
\setcounter{section}{0}

\section{Most general second order counterterms}
\label{appendix CT}

In what follows, we consider a free scalar field living on a
flat six-dimensional spacetime with a conical singularity at $r=0$.
As shown in \eqref{b-b prop} and \eqref{b-b prop 2}, the brane-brane
propagator of this field diverges logarithmically in the
thin-brane limit: $D_k(0,0)\sim \log \Lambda/k$ as $\Lambda\rightarrow
\infty$.
In order to make this quantity finite, one can try to include
counterterms both in the bulk and the brane. Unlike in usual
EFT, these counterterms are not added to make the interacting theory finite,
but the classical free theory itself.
We consider the following most general set of counterterms to
renormalize the free theory (renormalization of the wave function and mass both in the bulk and brane)
\ba
S=-\int \d ^6x \left[(1+Z_1)(\partial_a \phi)^2+\frac 12 M^2 \phi^2
+\delta^{(2)}(y)\(Z_2 (\partial \phi)^2+\frac 12 \lambda_2
\phi^2\)\right]\,.
\ea
The propagator will now be instead
\ba
D^\Lambda_k(r,r')\hspace{-10pt}&=&\hspace{-10pt}\sum_{n=-\infty}^{+\infty}\int_0 \frac{q \d
q}{2 \pi \alpha}\frac{e^{i \tilde n (\theta-\theta')}}{(1+Z_1(\Lambda))(q^2+k^2)+M^2(\Lambda)}J_{|\tilde n|}(q r)J_{|\tilde n|}(q
r')\nn\\
\hspace{-10pt}&=&\hspace{-10pt\sum_{n=-\infty}^{+\infty}
\Bigg[K_{|\tilde n|}\(\sqrt{k^2+\mu^2}\, r\)I_{|\tilde n|}\(\sqrt{k^2+\mu^2}\, r'\) \Theta (r-r')\nn\\
\label{D}
\hspace{-10pt}&&\hspace{-10pt}\phantom{\sum_{n=-\infty}^{+\infty}
\Bigg[}
+ (r\leftrightarrow r')
\Bigg]}\frac{e^{i \tilde n(\theta-\theta')}}{2\pi \alpha (1+Z_1)}\,,
\ea
where $\tilde n=n/\alpha$ and
$\mu^2(\Lambda)=\frac{M^2(\Lambda)}{1+Z_1(\Lambda)}$. Because of the
brane counterterms $Z_2$ and $\lambda_2$, this is however not
the complete two-point function. The two-point function is obtained
by summing over all the interactions with the bulk coupling. This
gives rise to following ``dressed" propagators
\ba
\hspace{-10pt}G_k(r,r')\hspace{-10pt}&=&\hspace{-10pt}D^\Lambda_k(r,r')-\frac{Z_2(\Lambda)
k^2+\lambda_2(\Lambda)}{1+(Z_2(\Lambda)k^2+\lambda_2(\Lambda))D_k^\Lambda(0,0)}D_k^\Lambda(r,0)D_k^\Lambda(0,r')\\
\hspace{-10pt}G_k(r,0)\hspace{-10pt}&=&\hspace{-10pt}D^\Lambda_k(r,0)\left[1-\frac{Z_2(\Lambda)
k^2+\lambda_2(\Lambda)}{1+(Z_2(\Lambda)k^2+\lambda_2(\Lambda))D_k^\Lambda(0,0)}D_k^\Lambda(0,0)\right]\\
\hspace{-10pt}G_k(0,0)\hspace{-10pt}&=&\hspace{-10pt}D^\Lambda_k(0,0)\left[1-\frac{Z_2(\Lambda)
k^2+\lambda_2(\Lambda)}{1+(Z_2(\Lambda)k^2+\lambda_2(\Lambda))D_k^\Lambda(0,0)}D_k^\Lambda(0,0)\right]\,.
\ea
Notice that in this approach, we no longer require the propagator $D^\Lambda_k(r,r')$ to be finite in the
thin brane limit but require instead that the two-point function $G_k$ between any
two points (taken in the bulk or the conical tip) is finite \ie $Z_1, Z_2, M^2$ and
$\lambda_2$ should flow in such a way that $G_k(r,r'), G_k(r,0)$ and
$G_k(0,0)$ are all finite. This implies that:
\begin{itemize}
\item $D_k^\Lambda(0,0)/D_k^\Lambda(r,0)$ should be finite in the limit $\Lambda\rightarrow\infty$ for any value
of $r$,
\item and similarly the quantity
$\left[D_k^\Lambda(r,r')-D_k^\Lambda(r,0)D_k^\Lambda(0,r')/D_k^\Lambda(0,0)\right]$
should be finite for any $r$ and $r'$.
\end{itemize}

It will therefore only be possible to make sense of the two-point
function on the brane, if one can renormalize the wave function and
the mass in such a way that the ratio
$D_k^\Lambda(0,0)/D_k^\Lambda(r,0)$ is finite. From \eqref{D}, we
get
\ba
D_k^\Lambda(r,0)=\frac{1}{2\pi \alpha (1+Z_1(\alpha))} K_0\(\sqrt{k^2+\mu^2(\Lambda)}\,
r\)\,,
\ea
and
\ba
D_k^\Lambda(0,0)=\lim_{\Lambda\rightarrow \infty}
D_k^\Lambda(\Lambda^{-1},0)=\frac{1}{2\pi \alpha
(1+Z_1(\alpha))} \log \frac{\Lambda}{\sqrt{k^2+\mu^2(\Lambda)}}\,.
\ea
It is therefore clear from these expressions that no matter how
the wave function and the mass renormalization flow, the quantity
$(D_k^\Lambda(r,0)/D_k^\Lambda(0,0))$ will never be finite in the
thin brane limit:
\ba
\frac{D_k^\Lambda(0,0)}{D_k^\Lambda(r,0)}=\frac{\log \Lambda-\frac12 \log (k^2+\mu^2(\Lambda))}{K_0 (\sqrt{k^2+\mu^2(\Lambda)}\,
r)} \rightarrow \infty\hspace{10pt}\text{ as }\ \Lambda\rightarrow
\infty,
\hspace{10pt}\forall \ \ \mu (\Lambda)\,.
\ea
No local counterterm (quadratic in the field) will thus ever make
the two-point function finite everywhere both in the bulk and the
brane.

\section{General $N$-point function}
\label{appendix generalcase}

In this appendix, we demonstrate that the RG flows of the brane
couplings $m$, $\beta_3$, $\beta_4$, $\lambda$, $\lambda_2$,
$\lambda_3$,
(\ref{RenormCouplings}, \ref{RGflow3}, \ref{beta3_2} and
\ref{beta4}) is sufficient to make all $N$-point Green's functions
finite in the thin-brane limit at any order in the loop expansion, hence justifying the renormalizability of
the theory. We focus, in what follows, on divergences associated with
the codimension-two nature of the theory and do not discuss loop
divergences which can be renormalized the standard way.

To simplify, we present the argument for the $N$-point Green's functions
having only brane field external legs $\chi$. As seen in section \ref{sectionInteractions},
the generalization to an arbitrary number of bulk field legs $\phi$ is
straight forward.

Defining the generating functional $G[J]$
\ba
G[J]=\int \mathcal{D}\left[\chi,\phi\right]
\langle 0|e^{-i(\mathcal H^{\rm int}_{\chi\phi} +J
\chi)}|0\rangle\,,
\ea
the $N$-point Green's functions are expressed by
\ba
G^{(N)}(x_1,\cdots,x_N)=\left.\frac{\delta^n G[J]}{\delta J(x_1)\cdots \delta
J(x_N)}\right|_{J=0}\,.
\ea
Technically, we are only interested in the connected Green's functions which
are generated by $G_c[J]=-i \log G[J]$. The connected $N$-point
Green's function can thus be expressed in terms of the lower ones as
\ba
G_c^{(N)}=-i\(\frac{G^{(N)}}{G^{(0)}}+\sum_{n=1}^{N-1}C^n_{N-1}\frac{G^{(n)}}{G^{(0)}}G_c^{(N-n)}\)\,,
\ea
where $C^n_{m}=\(\begin{array}{c}n\\m\end{array}\)$ is the  binomial
coefficient.

We have previously demonstrated that all the  connected tree-level Green's functions $G_c^{(n)}$
were finite for \nobreak{$n<\nobreak{5}$}.
In what follows, we show that this results remains true for any
Green's function $G^{(N)}$, $N\ge0$ at any order in the loop expansion, thus ensuring that all the connected
Green's is finite for any arbitrary number of
external legs.

The expression for the $N$-point function is
\ba
G^{(N)}(x_1,\cdots,x_N)&=&
\sum_{n\ge0}\frac{(-i)^n}{n!}\Big\langle
\chi(x_1)\cdots\chi(x_N) \nn\\
&& \times \prod_{i=1}^n\int \d y_i \(\beta_3 \chi^3(y_i) +\beta_4
\chi^4(y_i)+\lambda_3 \phi(0,y_i)\chi^2(y_i)\)\Big\rangle\nn\,.
\ea
Omitting the evaluation points $x_i$ and $y_i$ and remembering
that every field is evaluated on the brane, we have
\ba
G^{(N)}&=&\sum_{n\ge 0}\frac{(-i)^n}{n!}\ \big\langle\chi^N\(\beta_3 \chi^3 +\beta_4
\chi^4+\lambda_3 \phi\chi^2\)^n\big\rangle\nn\\
&=&\sum_{n\ge 0}\sum_{m=0}^n\sum_{k=0}^m\frac{(-i)^n}{n!}C^m_n C^k_m
\beta_3^{m-k}\beta_4^{n-m}
\lambda_3^k \ \Big\langle\chi^{N-k+4n-m}\phi^k \Big\rangle\nn
\ea
among these diagrams, some of them  can connect two bulk fields together, generating a singular
two-point function $G^{\phi\phi}(0,0)$. There can be $\alpha$ such
connections, (with $0\le\alpha\le k/2$), so that
\ba
G^{(N)}=\sum_{n\ge 0}\sum_{m=0}^n\sum_{k=0}^m\sum_{\alpha=0}^{k/2}\frac{(-i)^n}{n!}C^m_n C^k_m
C^{2\alpha}_k \beta_3^{m-k}\beta_4^{n-m}
\lambda_3^k \langle\chi^{N-k+4n-m}\phi^{k-2\alpha} \rangle \langle
\phi^{2\alpha}\rangle\,.\nn
\ea
We now consider the number of ways there is to connect the different
fields together (we recall that for now we are interested in all the
possible configurations, and do not restrict ourselves to the
connected ones). For $x$ fields $\langle\chi^x\rangle$, there
is\newline
$P_x=(x-1)(x-3)\cdots 3$ possible configurations if $x$ is even, and
no possible configurations if $x$ is odd ($\langle \chi\rangle=0$).
There is therefore $P_{2\alpha}$ ways to connect the $2\alpha$ bulk
fields in $\langle\phi^{2\alpha}\rangle$. To count the number of
configurations in $\langle\chi^{N-k+4n-m}\phi^{k-2\alpha} \rangle$, we need to pick first of all the fields $\chi$ which will
connect with the remaining $k-2\alpha$ fields $\phi$. There is
$(N-k+4n-m)!/(N-k+4n-m-(k-2\alpha))!$ such configurations and then
$P_{N-k+4n-m-(k-2\alpha)}$ ways to connect the remaining $\chi$
together.
Putting all this together, we therefore get
\ba
\label{G(N)1}
G^{(N)}\hspace{-5pt}&=&\hspace{-5pt}\sum_{n\ge
0}\sum_{m=0}^n\sum_{k=0}^m\sum_{\alpha=0}^{k/2}\Bigg(
\frac{(-i)^n}{n!}C^m_n C^k_m \  \beta_3^{m-k}\beta_4^{n-m}\lambda_3^k
C^{2\alpha}_k\\
&& \hspace{20pt}\times \ \frac{(N-k+4n-m)!}{(N-2k+4n-m+2\alpha)!}P_{N-2k+4n-m+2\alpha}P_{2\alpha}\nn \\
&& \hspace{20pt}\times \
\(G^{\chi\chi}\)^{\frac{N-2k+4n-m+2\alpha}{2}}\(G^{\chi\phi}(0)\)^{k-2\alpha}\(G^{\phi\phi}(0,0)\)^\alpha \Bigg)\nn
\ea
with
\ba
P_x=\left\{
\begin{array}{cl}
(x-1)(x-3)\cdots 3=\frac{2n!}{2^n n!} & \text{if } x \text{ is even},\ x=2n>0,\\
1 &  \text{if } x=0,\\
0 & \text{otherwise.}
\end{array}\right.
\ea
We therefore notice that in the previous expression \eqref{G(N)1} of
the Green's function, $N+m$ needs to be even.

Expressing the bulk-bulk and bulk-brane propagator as
\ba
G^{\chi\phi}(0)&=&-i\lambda \tilde D G^{\chi\chi}\nn\\
G^{\phi\phi}(0,0)&=&\tilde D-\lambda^2 \tilde D ^2
G^{\chi\chi}\nn\,,
\ea
we get
\ba
G^{(N)}\hspace{-5pt}&=&\hspace{-5pt}\sum_{n\ge
0}\sum_{m=0}^n\sum_{k=0}^m\sum_{\alpha=0}^{k/2}\sum_{\gamma=0}^\alpha
\frac{(-i)^n}{n!}C^m_n C^k_m
\beta_4^{n-m}\beta_3^{m-k}\lambda_3^k\lambda^{k-2\gamma}\tilde
D^{k-2\gamma} \(G^{\chi\chi}\)^{\frac{1}{2}(N+4n-m-2\gamma)}\nn \\
&&\hspace{-20pt}\times
\underbrace{(-1)^{\alpha-\gamma}(-i)^{k-2\alpha}
C^{2\alpha}_k C^\gamma_\alpha
\frac{(N-k+4n-m)!}{(N-2k+4n-m+2\alpha)!}P_{N-2k+4n-m+2\alpha}P_{2\alpha}}_{=F(\alpha)}\,.
\ea

Notice that the coefficient $\alpha$ does not affect the couplings,
and so the summation over $\alpha$ can be performed without
affecting the order of the diagram
\ba
\hspace{-10pt}\sum_{\alpha=0}^{k/2}\sum_{\gamma=0}^\alpha F(\alpha)=\sum_{\gamma=0}^{k/2} \sum_{\alpha=\gamma}^{k/2}
F(\alpha)
=\sum_{\gamma=0}^{k/2}(-i)^{k-2\gamma}\frac{2^\gamma
k!}{\gamma!(k-2\gamma)!}P_{N-m+4n+2\gamma}\,.
\ea
We now change the summation variables to $(n,m,k,\gamma)\rightarrow
(X,Y,l,\gamma)$, with
\ba
X=m-2\gamma\,,\hspace{10pt}Y=n-m+\gamma\,,\hspace{5pt}\text{and}\hspace{9pt}l=k-2\gamma
\ea
so that the Green's function can be expressed as
\ba
\hspace{-5pt}G^{(N)}\hspace{-5pt}&=&\hspace{-5pt}\sum_{X\ge0}\sum_{Y\ge0}\Bigg(
\frac{(-i)^{X+Y}}{X!Y!}P_{N+3X+4Y}  \(G^{\chi\chi}\)^{\frac{1}{2}(N+3X+4Y)} \nn \\
&& \hspace{40pt}\times \sum_{l=0}^X\sum_{\gamma=0}^Y  C^l_X C^\gamma_Y
\frac{(-i)^{l+\gamma}}{2^\gamma}\lambda^l
\lambda_3^{l+2\gamma}\beta_3^{X-l}\beta_4^{Y-\gamma}\tilde D
^{l+\gamma}\Bigg)\,,
\ea
Finally, summing over $l$ and $\gamma$, we recover the familiar
expressions
\ba
\hspace{-20pt}G^{(N)}=\sum_{X\ge0}\sum_{Y\ge0}
\frac{(-i)^{X+Y}}{X!Y!}P_{N+3X+4Y}
\(G^{\chi\chi}\)^{\frac{1}{2}(N+3X+4Y)}
\(\beta_3-i \lambda \lambda_3 \tilde D\)^X\, \(\beta_4-\frac i 2 \lambda_3^2 \tilde
D\)^Y\,.
\ea
So the coupling constants $\lambda$, $\beta_3$, $\beta_3$, $\lambda_3$ and the free bulk-bulk
propagator $\tilde D$ come in
precisely the right combination to be finite. The RG flows of
$\lambda_3$, $\beta_3$ and $\beta_4$ indeed ensures that both
expressions $\(\beta_3-i \lambda \lambda_3 \tilde D\)$ and
$ \(\beta_4-\frac i 2 \lambda_3^2 \tilde D\)$ are finite, see
eqs.~\eqref{3ptfinite} and \eqref{beta4}.
Since the
renormalization of $\lambda$, $m^2$ and $\lambda_2$ is such that the
brane propagator $G^{\chi\chi}$ is finite, we can conclude that
$G^{(N)}$ is completely finite in the thin brane limit (up to loop
diagram divergences which can be renormalized in the standard way).
This argument thus demonstrates that the theory is renormalizable.


\begin{thebibliography}{99}%{is-unsrt}


%\cite{Randall:1999ee}
\bibitem{Randall:1999ee}
  L.~Randall and R.~Sundrum,
  ``A large mass hierarchy from a small extra dimension,''
  Phys.\ Rev.\ Lett.\  {\bf 83}, 3370 (1999)
  [arXiv:hep-ph/9905221];
  %%CITATION = PRLTA,83,3370;%%
%
 L.~Randall and R.~Sundrum,
  ``An alternative to compactification,''
  Phys.\ Rev.\ Lett.\  {\bf 83}, 4690 (1999)
  [arXiv:hep-th/9906064].
  %%CITATION = PRLTA,83,4690;%%



%
%\cite{Salam:1984cj}
\bibitem{Salam:1984cj}
  A.~Salam and E.~Sezgin,
   ``Chiral Compactification On Minkowski X S**2 Of N=2 Einstein-Maxwell
  Supergravity In Six-Dimensions,''
  Phys.\ Lett.\  B {\bf 147}, 47 (1984).
  %%CITATION = PHLTA,B147,47;%%

 %\cite{Kanti:2001vb}
\bibitem{Kanti:2001vb}
  P.~Kanti, R.~Madden and K.~A.~Olive,
  ``A 6-D brane world model,''
  Phys.\ Rev.\ D {\bf 64}, 044021 (2001)
  [arXiv:hep-th/0104177];
  %%CITATION = HEP-TH 0104177;%%
%\cite{Gibbons:2003di}
%\bibitem{Gibbons:2003di}
  G.~W.~Gibbons, R.~Guven and C.~N.~Pope,
  ``3-branes and uniqueness of the Salam-Sezgin vacuum,''
  Phys.\ Lett.\  B {\bf 595}, 498 (2004)
  [arXiv:hep-th/0307238];
  %%CITATION = PHLTA,B595,498;%%
%\cite{Gibbons:2003di}
%\cite{Aghababaie:2003ar}
%\bibitem{Aghababaie:2003ar}
  Y.~Aghababaie {\it et al.},
  ``Warped brane worlds in six dimensional supergravity,''
  JHEP {\bf 0309}, 037 (2003)
  [arXiv:hep-th/0308064];
  %%CITATION = HEP-TH 0308064;%%
  %\cite{Lee:2004vn}
%\bibitem{Lee:2004vn}
  H.~M.~Lee and A.~Papazoglou,
  ``Brane solutions of a spherical sigma model in six dimensions,''
  Nucl.\ Phys.\  B {\bf 705}, 152 (2005)
  [arXiv:hep-th/0407208];
  %%CITATION = NUPHA,B705,152;%%
  %\cite{Burgess:2004dh}
%\bibitem{Burgess:2004dh}
  C.~P.~Burgess, F.~Quevedo, G.~Tasinato and I.~Zavala,
   ``General axisymmetric solutions and self-tuning in 6D chiral gauged
  supergravity,''
  JHEP {\bf 0411}, 069 (2004)
  [arXiv:hep-th/0408109];
  %%CITATION = JHEPA,0411,069;%%
%\cite{Mukohyama:2005yw}
%\bibitem{Mukohyama:2005yw}
  S.~Mukohyama, Y.~Sendouda, H.~Yoshiguchi and S.~Kinoshita,
  ``Warped flux compactification and brane gravity,''
  JCAP {\bf 0507}, 013 (2005)
  [arXiv:hep-th/0506050];
  %%CITATION = HEP-TH 0506050;%%
%
%\cite{Parameswaran:2005mm}
%\bibitem{Parameswaran:2005mm}
  S.~L.~Parameswaran, G.~Tasinato and I.~Zavala,
  ``The 6D SuperSwirl,''
  Nucl.\ Phys.\  B {\bf 737}, 49 (2006)
  [arXiv:hep-th/0509061];
  %%CITATION = NUPHA,B737,49;%%
%
%\cite{Lee:2005az}
% \bibitem{Lee:2005az}
  H.~M.~Lee and C.~Ludeling,
   ``The general warped solution with conical branes in six-dimensional
  supergravity,''
  JHEP {\bf 0601}, 062 (2006)
  [arXiv:hep-th/0510026];
  %%CITATION = JHEPA,0601,062;%%
 A.~J.~Tolley, C.~P.~Burgess, D.~Hoover and Y.~Aghababaie,
  ``Bulk singularities and the effective cosmological constant for higher
  %co-dimension branes,''
  JHEP {\bf 0603}, 091 (2006)
  [arXiv:hep-th/0512218];
  %
  %\cite{Tolley:2006ht}
%\bibitem{Tolley:2006ht}
  A.~J.~Tolley, C.~P.~Burgess, C.~de Rham and D.~Hoover,
  ``Scaling solutions to 6D gauged chiral supergravity,''
  New J.\ Phys.\  {\bf 8}, 324 (2006)
  [arXiv:hep-th/0608083];
  %%CITATION = NJOPF,8,324;%%
  %\cite{Kobayashi:2007hf}
% \bibitem{Kobayashi:2007hf}
  T.~Kobayashi and M.~Minamitsuji,
  ``Brane cosmological solutions in six-dimensional warped flux
  compactifications,''
  arXiv:0705.3500 [hep-th];
  %%CITATION = ARXIV:0705.3500;%%
 E.~J.~Copeland and O.~Seto,
  ``Dynamical solutions of warped six dimensional supergravity,''
  arXiv:0705.4169 [hep-th].
  %%CITATION = ARXIV:0705.4169;%%

%%7. cosmology of codimension 2 branes
%
%\cite{Hayakawa:2003qm}
\bibitem{Hayakawa:2003qm}
  S.~Hayakawa, D.~Ida, T.~Shiromizu and T.~Tanaka,
  ``Gravitation in the codimension two brane world,''
  Prog.\ Theor.\ Phys.\ Suppl.\  {\bf 148}, 128 (2003);
  %%CITATION = PTPSA,148,128;%%
%
%\cite{Cline:2003ak}
%\bibitem{Cline:2003ak}
 J.~M.~Cline, J.~Descheneau, M.~Giovannini and J.~Vinet,
 ``Cosmology of codimension-two braneworlds,''
 JHEP {\bf 0306}, 048 (2003)
 [arXiv:hep-th/0304147];
%%CITATION = HEP-TH 0304147;%%
%
%\cite{Vinet:2004dc}
%\bibitem{Vinet:2004dc}
  J.~Vinet,
  ``Generalised cosmology of codimension-two braneworlds,''
  Int.\ J.\ Mod.\ Phys.\ A {\bf 19}, 5295 (2004)
  [arXiv:hep-th/0408082];
  %%CITATION = HEP-TH 0408082;%%
%
%\cite{Charmousis:2005ez}
%\bibitem{Charmousis:2005ez}
 C.~Charmousis and R.~Zegers,
 ``Einstein gravity on an even codimension brane,''
 Phys.\ Rev.\ D {\bf 72}, 064005 (2005)
 [arXiv:hep-th/0502171];
 %%CITATION = HEP-TH 0502171;%%
%
%\cite{Papantonopoulos:2005nw}
%\bibitem{Papantonopoulos:2005nw}
 E.~Papantonopoulos and A.~Papazoglou,
 ``Cosmological evolution of a purely conical codimension-2 brane world,''
 JHEP {\bf 0509}, 012 (2005)
 [arXiv:hep-th/0507278].
 %%CITATION = HEP-TH 0507278;%%
%
%%\cite{Gibbons:1986wg}
%\bibitem{Gibbons:1986wg}
%  G.~W.~Gibbons and D.~L.~Wiltshire,
%  %``Space-Time As A Membrane In Higher Dimensions,''
%  Nucl.\ Phys.\ B {\bf 287} (1987) 717
%  [arXiv:hep-th/0109093].
%  %%CITATION = HEP-TH 0109093;%%
%  %%Cited 78 times in SPIRES-HEP
%
% %%%%%%%%%%%%%%%%%%%%%%%%%%%%%%%%%%%%%%%%%%%%%%%%%%%%%%%%%%%%%%%%%%%%%%%%%%%%%%%%%%%%%%%%%%%%%%%%%%%%%%
% Stability




%\cite{Graesser:2004xv}
\bibitem{Graesser:2004xv}
  M.~L.~Graesser, J.~E.~Kile and P.~Wang,
  ``Gravitational perturbations of a six dimensional self-tuning model,''
  Phys.\ Rev.\ D {\bf 70}, 024008 (2004)
  [arXiv:hep-th/0403074];
  %%CITATION = HEP-TH 0403074;%%
%\cite{Lee:2006ge}
%\bibitem{Lee:2006ge}
  H.~M.~Lee and A.~Papazoglou,
  ``Scalar mode analysis of the warped Salam-Sezgin model,''
  Nucl.\ Phys.\  B {\bf 747}, 294 (2006)
  [Erratum-ibid.\  B {\bf 765}, 200 (2007)]
  [arXiv:hep-th/0602208];
  %%CITATION = NUPHA,B747,294;%%
%\cite{Burgess:2006ds}
%\bibitem{Burgess:2006ds}
  C.~P.~Burgess, C.~de Rham, D.~Hoover, D.~Mason and A.~J.~Tolley,
  ``Kicking the rugby ball: Perturbations of 6D gauged chiral supergravity,''
  JCAP {\bf 0702}, 009 (2007)
  [arXiv:hep-th/0610078];
  %%CITATION = JCAPA,0702,009;%%
%
%\cite{Parameswaran:2007cb}
% \bibitem{Parameswaran:2007cb}
  S.~L.~Parameswaran, S.~Randjbar-Daemi and A.~Salvio,
  ``Stability and Negative Tensions in 6D Brane Worlds,''
  arXiv:0706.1893 [hep-th].
  %%CITATION = ARXIV:0706.1893;%%



%\cite{Peloso:2006cq}
\bibitem{Peloso:2006cq}
  M.~Peloso, L.~Sorbo and G.~Tasinato,
  ``Standard 4d gravity on a brane in six dimensional flux compactifications,''
  Phys.\ Rev.\  D {\bf 73}, 104025 (2006)
  [arXiv:hep-th/0603026].
  %%CITATION = PHRVA,D73,104025;%%

  %\cite{Arkani-Hamed:1998rs}
\bibitem{Arkani-Hamed:1998rs}
  N.~Arkani-Hamed, S.~Dimopoulos and G.~R.~Dvali,
  ``The hierarchy problem and new dimensions at a millimeter,''
  Phys.\ Lett.\ B {\bf 429}, 263 (1998)
  [arXiv:hep-ph/9803315].
  %%CITATION = HEP-PH 9803315;%%

%\cite{Albrecht:2001xt}
\bibitem{Albrecht:2001xt}
  A.~Albrecht, C.~P.~Burgess, F.~Ravndal and C.~Skordis,
  ``Natural quintessence and large extra dimensions,''
  Phys.\ Rev.\  D {\bf 65}, 123507 (2002)
  [arXiv:astro-ph/0107573];
  %%CITATION = PHRVA,D65,123507;%%
%\cite{Ghilencea:2005vm}
%\bibitem{Ghilencea:2005vm}
  D.~M.~Ghilencea, D.~Hoover, C.~P.~Burgess and F.~Quevedo,
   ``Casimir energies for 6D supergravities compactified on T(2)/Z(N) with
  Wilson lines,''
  JHEP {\bf 0509}, 050 (2005)
  [arXiv:hep-th/0506164];
  %%CITATION = JHEPA,0509,050;%%
  %\cite{Elizalde:2007di}
% \bibitem{Elizalde:2007di}
  E.~Elizalde, M.~Minamitsuji and W.~Naylor,
  ``Casimir effect in rugby-ball type flux compactifications,''
  Phys.\ Rev.\  D {\bf 75}, 064032 (2007)
  [arXiv:hep-th/0702098];
  %%CITATION = PHRVA,D75,064032;%%
%\cite{Minamitsuji:2007iz}
% \bibitem{Minamitsuji:2007iz}
  M.~Minamitsuji,
  ``Casimir effect in a 6D warped flux compactification model,''
  arXiv:0704.3623 [gr-qc];
  %%CITATION = ARXIV:0704.3623;%%
%
%\cite{Koyama:2007rx}
%\bibitem{Koyama:2007rx}
  K.~Koyama,
  ``The cosmological constant and dark energy
  in braneworlds,''
  arXiv:0706.1557 [astro-ph].
  %%CITATION = ARXIV:0706.1557;%%


%
%% codimension2 branes in 6d and the cosmological constant problem
%

%\cite{Rubakov:1983bz}
\bibitem{Rubakov:1983bz}
 V.~A.~Rubakov and M.~E.~Shaposhnikov,
 ``Extra Space-Time Dimensions: Towards A Solution To The Cosmological
 Constant Problem,''
 Phys.\ Lett.\ B {\bf 125}, 139 (1983).
  %%CITATION = PHLTA,B125,139;%%

%\cite{Aghababaie:2003wz}
 \bibitem{Aghababaie:2003wz}
 Y.~Aghababaie, C.~P.~Burgess, S.~L.~Parameswaran and F.~Quevedo,
 ``Towards a naturally small cosmological constant from branes in 6D
 supergravity,''
 Nucl.\ Phys.\ B {\bf 680}, 389 (2004)
 [arXiv:hep-th/0304256];
 %%CITATION = HEP-TH 0304256;%%
%
 %\cite{Corradini:2002es}
%\bibitem{Corradini:2002es}
  O.~Corradini, A.~Iglesias and Z.~Kakushadze,
  ``Toward solving the cosmological constant problem?,''
  Int.\ J.\ Mod.\ Phys.\ A {\bf 18}, 3221 (2003)
  [arXiv:hep-th/0212101];
  %%CITATION = HEP-TH 0212101;%%
%
%\cite{Carroll:2003db}
% \bibitem{Carroll:2003db}
S.~M.~Carroll and M.~M.~Guica,
``Sidestepping the cosmological constant with football-shaped extra
dimensions,''
arXiv:hep-th/0302067;
%%CITATION = HEP-TH 0302067;%%
%
%
%\cite{Navarro:2003vw}
%\bibitem{Navarro:2003vw}
 I.~Navarro,
 ``Codimension two compactifications and the cosmological constant  problem,''
 JCAP {\bf 0309}, 004 (2003)
 [arXiv:hep-th/0302129];
 %%CITATION = HEP-TH 0302129;%%
%
%
%\cite{Navarro:2003bf}
% \bibitem{Navarro:2003bf}
 I.~Navarro,
 ``Spheres, deficit angles and the cosmological constant,''
 Class.\ Quant.\ Grav.\  {\bf 20}, 3603 (2003)
 [arXiv:hep-th/0305014];
 %%CITATION = HEP-TH 0305014;%%
%
%
%\cite{Nilles:2003km}
% \bibitem{Nilles:2003km}
  H.~P.~Nilles, A.~Papazoglou and G.~Tasinato,
  ``Selftuning and its footprints,''
  Nucl.\ Phys.\ B {\bf 677}, 405 (2004)
  [arXiv:hep-th/0309042];
%%CITATION = HEP-TH 0309042;%%
%
%\cite{Burgess:2004kd}
%\bibitem{Burgess:2004kd}
  C.~P.~Burgess,
   ``Supersymmetric large extra dimensions and the cosmological constant: An
  update,''
  Annals Phys.\  {\bf 313}, 283 (2004)
  [arXiv:hep-th/0402200];
  %%CITATION = APNYA,313,283;%%
%\cite{Schwindt:2005ns}
%\bibitem{Schwindt:2005ns}
 J.~M.~Schwindt and C.~Wetterich,
 ``The cosmological constant problem in codimension-two brane models,''
 Phys.\ Lett.\ B {\bf 628}, 189 (2005)
 [arXiv:hep-th/0508065];
 %%CITATION = HEP-TH 0508065;%%
%
%\cite{Burgess:2005wu}
% \bibitem{Burgess:2005wu}
 C.~P.~Burgess,
 ``Supersymmetric large extra dimensions and the cosmological constant
 problem,''
 arXiv:hep-th/0510123;
 %%CITATION = HEP-TH 0510123;%%
%\cite{Park:2007ij}
%\bibitem{Park:2007ij}
  E.~K.~Park and P.~S.~Kwon,
  ``A self-tuning mechanism in 6d gravity-scalar theory,''
  arXiv:hep-th/0702171.
  %%CITATION = HEP-TH/0702171;%%


 %\cite{Burgess:2004yq}
\bibitem{Burgess:2004yq}
  C.~P.~Burgess, J.~Matias and F.~Quevedo,
  ``MSLED: A minimal supersymmetric large extra dimensions scenario,''
  Nucl.\ Phys.\  B {\bf 706}, 71 (2005)
  [arXiv:hep-ph/0404135];
  %%CITATION = NUPHA,B706,71;%%
%
%\cite{Vinet:2004bk}
% \bibitem{Vinet:2004bk}
 J.~Vinet and J.~M.~Cline,
 ``Can codimension-two branes solve the cosmological constant problem?,''
 Phys.\ Rev.\ D {\bf 70}, 083514 (2004)
 [arXiv:hep-th/0406141];
 %%CITATION = HEP-TH 0406141;%%
%
%\cite{Garriga:2004tq}
%\bibitem{Garriga:2004tq}
 J.~Garriga and M.~Porrati,
 ``Football shaped extra dimensions and the absence of self-tuning,''
 JHEP {\bf 0408}, 028 (2004)
 [arXiv:hep-th/0406158];
 %%CITATION = HEP-TH 0406158;%%
%
%\cite{Vinet:2005dg}
%\bibitem{Vinet:2005dg}
 J.~Vinet and J.~M.~Cline,
 ``Codimension-two branes in six-dimensional supergravity and the cosmological
 constant problem,''
 Phys.\ Rev.\ D {\bf 71}, 064011 (2005)
 [arXiv:hep-th/0501098].
 %%CITATION = HEP-TH 0501098;%%


%\cite{Dvali:2002pe}
 \bibitem{Dvali:2002pe}
  G.~Dvali, G.~Gabadadze and M.~Shifman,
  ``Diluting cosmological constant in infinite volume extra dimensions,''
  Phys.\ Rev.\ D {\bf 67}, 044020 (2003)
  [arXiv:hep-th/0202174];
  %%CITATION = HEP-TH 0202174;%%
 %\cite{Corradini:2003qp}
%\bibitem{Corradini:2003qp}
  O.~Corradini, A.~Iglesias and Z.~Kakushadze,
  ``Diluting solutions of the cosmological constant problem,''
  Mod.\ Phys.\ Lett.\ A {\bf 18}, 1343 (2003)
  [arXiv:hep-th/0305164].
  %%CITATION = HEP-TH 0305164;%%

%%%%%%%%%%%%%%%%%%%%%%%%%%%%%%%%%%%%%%%%%%%%%%%%%%%%%%%%
%%%%   DGP model

%\cite{Dvali:2007kt}
\bibitem{Dvali:2007kt}
  G.~Dvali, S.~Hofmann and J.~Khoury,
  %``Degravitation of the cosmological constant and graviton width,''
  arXiv:hep-th/0703027.
  %%CITATION = HEP-TH/0703027;%%



%\cite{Dvali:2000rv}
\bibitem{Dvali:2000rv}
  G.~R.~Dvali, G.~Gabadadze and M.~Porrati,
  ``Metastable gravitons and infinite volume extra dimensions,''
  Phys.\ Lett.\  B {\bf 484}, 112 (2000)
  [arXiv:hep-th/0002190];
  %%CITATION = PHLTA,B484,112;%%
%\cite{Dvali:2000hr}
%\bibitem{Dvali:2000hr}
  G.~R.~Dvali, G.~Gabadadze and M.~Porrati,
  ``4D gravity on a brane in 5D Minkowski space,''
  Phys.\ Lett.\  B {\bf 485}, 208 (2000)
  [arXiv:hep-th/0005016];
    %%CITATION = PHLTA,B485,208;%%
%\cite{Dvali:2000xg}
%\bibitem{Dvali:2000xg}
  G.~R.~Dvali and G.~Gabadadze,
  ``Gravity on a brane in infinite-volume extra space,''
  Phys.\ Rev.\  D {\bf 63}, 065007 (2001)
  [arXiv:hep-th/0008054].
  %%CITATION = PHRVA,D63,065007;%%



%\cite{Weinberg:1988cp}
\bibitem{Weinberg:1988cp}
  S.~Weinberg,
  ``The cosmological constant problem,''
  Rev.\ Mod.\ Phys.\  {\bf 61}, 1 (1989).
  %%CITATION = RMPHA,61,1;%%

  %
%%%%%%%%%%%%%%%%%%%%%%%%%%%%%%%%%%%%%%%%%%%%%%%%%%%%%%%%%%%%%%%%%%%%%%%%%%%%%%%%%%%%%%%%%%%%%%%%%%%%%%%
%%%%% Geroch and Traschen


  %\cite{Geroch:1987qn}
\bibitem{Geroch:1987qn}
  R.~Geroch and J.~H.~Traschen,
  ``Strings And Other Distributional Sources In General Relativity,''
  Phys.\ Rev.\ D {\bf 36}, 1017 (1987).
  %%CITATION = PHRVA,D36,1017;%%


%\cite{Goldberger:2001tn}
\bibitem{Goldberger:2001tn}
  W.~D.~Goldberger and M.~B.~Wise,
  ``Renormalization group flows for brane couplings,''
  Phys.\ Rev.\ D {\bf 65}, 025011 (2002)
  [arXiv:hep-th/0104170].
  %%CITATION = HEP-TH 0104170;%%
  %%Cited 25 times in SPIRES-HEP

%%%%%%%%%%%%%%%%%%%%%%%%%%%%%%%%%%%%%%%%%%%%%%%%%%%%%%%
% Goldberger-Wise approach

%\cite{Goldberger:2004jt}
\bibitem{Goldberger:2004jt}
  W.~D.~Goldberger and I.~Z.~Rothstein,
  ``An effective field theory of gravity for extended objects,''
  Phys.\ Rev.\  D {\bf 73}, 104029 (2006)
  [arXiv:hep-th/0409156];
  %%CITATION = PHRVA,D73,104029;%%
%
%\cite{Goldberger:2005cd}
% \bibitem{Goldberger:2005cd}
  W.~D.~Goldberger and I.~Z.~Rothstein,
  ``Dissipative effects in the worldline approach to black hole dynamics,''
  Phys.\ Rev.\  D {\bf 73}, 104030 (2006)
  [arXiv:hep-th/0511133];
  %%CITATION = PHRVA,D73,104030;%%
%\cite{Goldberger:2006bd}
% \bibitem{Goldberger:2006bd}
  W.~D.~Goldberger and I.~Z.~Rothstein,
  ``Towers of gravitational theories,''
  Gen.\ Rel.\ Grav.\  {\bf 38}, 1537 (2006)
  [Int.\ J.\ Mod.\ Phys.\  D {\bf 15}, 2293 (2006)]
  [arXiv:hep-th/0605238];
  %%CITATION = IMPAE,D15,2293;%%
%\cite{Goldberger:2007hy}
%\bibitem{Goldberger:2007hy}
  W.~D.~Goldberger,
  ``Les Houches lectures on effective field theories and gravitational
  radiation,''
  arXiv:hep-ph/0701129.
  %%CITATION = HEP-PH/0701129;%%


%\cite{Porto:2005ac}
\bibitem{Porto:2005ac}
  R.~A.~Porto,
  ``Post-Newtonian corrections to the motion of spinning bodies in NRGR,''
  Phys.\ Rev.\  D {\bf 73}, 104031 (2006)
  [arXiv:gr-qc/0511061];
  %%CITATION = PHRVA,D73,104031;%%
%\cite{Porto:2006bt}
% \bibitem{Porto:2006bt}
  R.~A.~Porto and I.~Z.~Rothstein,
  ``The hyperfine Einstein-Infeld-Hoffmann potential,''
  Phys.\ Rev.\ Lett.\  {\bf 97}, 021101 (2006)
  [arXiv:gr-qc/0604099];
  %%CITATION = PRLTA,97,021101;%%
%\cite{Porto:2007pw}
% \bibitem{Porto:2007pw}
  R.~A.~Porto and R.~Sturani,
  ``Scalar gravity: Post-Newtonian corrections via an effective field theory
  approach,''
  arXiv:gr-qc/0701105;
  %%CITATION = GR-QC/0701105;%%
%\cite{Porto:2007px}
%\bibitem{Porto:2007px}
  R.~A.~Porto,
  ``New results at 3PN via an effective field theory of gravity,''
  arXiv:gr-qc/0701106.
  %%CITATION = GR-QC/0701106;%%


%%%%%%%%%%%%%%%%%%%%%%%%%%%%%%%%%%%%%%%%%%%%%%%%
%%%% BRANE LOCALISED KINETIC TERMS

%\cite{del Aguila:2003bh}
\bibitem{del Aguila:2003bh}
  F.~del Aguila, M.~Perez-Victoria and J.~Santiago,
  ``Bulk fields with general brane kinetic terms,''
  JHEP {\bf 0302}, 051 (2003)
  [arXiv:hep-th/0302023];
  %%CITATION = JHEPA,0302,051;%%
%
%\cite{del Aguila:2006kj}
%\bibitem{del Aguila:2006kj}
  F.~del Aguila, M.~Perez-Victoria and J.~Santiago,
  ``Effective description of brane terms in extra dimensions,''
  JHEP {\bf 0610}, 056 (2006)
  [arXiv:hep-ph/0601222].
  %%CITATION = JHEPA,0610,056;%%



%\cite{Corradini:2002ta}
\bibitem{Corradini:2002ta}
  O.~Corradini, A.~Iglesias, Z.~Kakushadze and P.~Langfelder,
  ``A remark on smoothing out higher codimension branes,''
  Mod.\ Phys.\ Lett.\ A {\bf 17}, 795 (2002)
  [arXiv:hep-th/0201201].
  %%CITATION = HEP-TH 0201201;%%



%%%%%%%%%%%%%%%%%%%%%%%%%%%%%%%%%%%%%%%%%%%%%%%%%%%%%
% Different regularizations

%\cite{Gherghetta:2000qi}
\bibitem{Gherghetta:2000qi}
 T.~Gherghetta and M.~E.~Shaposhnikov,
 ``Localizing gravity on a string-like defect in six dimensions,''
 Phys.\ Rev.\ Lett.\  {\bf 85}, 240 (2000)
 [arXiv:hep-th/0004014].
 %%CITATION = HEP-TH 0004014;%%

%\cite{Navarro:2004di}
\bibitem{Navarro:2004di}
  I.~Navarro and J.~Santiago,
  ``Gravity on codimension 2 brane worlds,''
  JHEP {\bf 0502}, 007 (2005)
  [arXiv:hep-th/0411250];
  %%CITATION = HEP-TH 0411250;%%
%
%\cite{deRham:2005ci}
% \bibitem{deRham:2005ci}
  C.~de Rham and A.~J.~Tolley,
  %``Gravitational waves in a codimension two braneworld,''
  JCAP {\bf 0602}, 003 (2006)
  [arXiv:hep-th/0511138].
  %%CITATION = JCAPA,0602,003;%%


%\cite{Burgess:2007vi}
\bibitem{Burgess:2007vi}
%\cite{Kobayashi:2007kv}
%\bibitem{Kobayashi:2007kv}
  T.~Kobayashi and M.~Minamitsuji,
  ``Gravity on an extended brane in six-dimensional warped flux
  compactifications,''
  Phys.\ Rev.\  D {\bf 75}, 104013 (2007)
  [arXiv:hep-th/0703029];
  %%CITATION = PHRVA,D75,104013;%%
%
  C.~P.~Burgess, D.~Hoover and G.~Tasinato,
  ``UV Caps and Modulus Stabilization for 6D Gauged Chiral Supergravity,''
  arXiv:0705.3212 [hep-th].
  %%CITATION = ARXIV:0705.3212;%%



%\cite{Nelson:1999wu}
\bibitem{Nelson:1999wu}
 A.~E.~Nelson,
 ``A new angle on intersecting branes in infinite extra dimensions,''
 Phys.\ Rev.\ D {\bf 63}, 087503 (2001)
 [arXiv:hep-th/9909001];
 %%CITATION = HEP-TH 9909001;%%
 %
%
%
%\cite{Gravanis:2003aq}
%\bibitem{Gravanis:2003aq}
 E.~Gravanis and S.~Willison,
 ``Intersecting hyper-surfaces in dimensionally continued topological  density
 gravitation,''
 J.\ Math.\ Phys.\  {\bf 45}, 4223 (2004)
 [arXiv:hep-th/0306220];
 %%CITATION = HEP-TH 0306220;%%
%
%
%\cite{Navarro:2004kh}
%\bibitem{Navarro:2004kh}
 I.~Navarro and J.~Santiago,
 ``Higher codimension braneworlds from intersecting branes,''
 JHEP {\bf 0404}, 062 (2004)
 [arXiv:hep-th/0402204];
 %%CITATION = HEP-TH 0402204;%%
%
%\cite{Kaloper:2004cy}
%\bibitem{Kaloper:2004cy}
 N.~Kaloper,
 ``Origami world,''
 JHEP {\bf 0405}, 061 (2004)
 [arXiv:hep-th/0403208].
 %%CITATION = HEP-TH 0403208;%%



%\cite{Cod2Cod1}
\bibitem{Cod2Cod1}
  C.~de Rham, G.~Dvali, S.~Hofmann, J.~Khoury, O.~Pujolas, M.~Redi and A.~J.~Tolley,
  ``Cascading DGP,''
  arXiv:0711.2072 [hep-th];
  %%CITATION = ARXIV:0711.2072;%%
  C.~de Rham, S.~Hofmann, J.~Khoury and A.~J.~Tolley,
  ``Cascading Gravity and Degravitation,''
  arXiv:0712.2821 [hep-th].
  %%CITATION = ARXIV:0712.2821;%%








%\cite{Weinberg:1965nx}
\bibitem{Weinberg:1965nx}
  S.~Weinberg,
  ``Infrared photons and gravitons,''
  Phys.\ Rev.\  {\bf 140}, B516 (1965).
  %%CITATION = PHRVA,140,B516;%%


%%%%%%%%%%%%%%%%%%%%%%%%%%%%%%%%%%%%%%%%%%%%%%%%%%%%%%%%%
% Localized Ricci term on the brane



%\cite{deRham:2006hs}
\bibitem{deRham:2006hs}
  C.~de Rham, T.~Shiromizu and A.~J.~Tolley,
  ``Weaker gravity at submillimetre scales in braneworlds models,''
  arXiv:gr-qc/0604071.
  %%CITATION = GR-QC/0604071;%%

%\cite{Deffayet:2001pu}
\bibitem{Deffayet:2001pu}
  C.~Deffayet, G.~R.~Dvali and G.~Gabadadze,
  ``Accelerated universe from gravity leaking to extra dimensions,''
  Phys.\ Rev.\  D {\bf 65}, 044023 (2002)
  [arXiv:astro-ph/0105068].
  %%CITATION = PHRVA,D65,044023;%%



%\cite{deRham:2006pe}
\bibitem{deRham:2006pe}
  C.~de Rham and A.~J.~Tolley,
  ``Mimicking Lambda with a spin-two ghost condensate,''
  JCAP {\bf 0607}, 004 (2006)
  [arXiv:hep-th/0605122];
  %%CITATION = JCAPA,0607,004;%%
  %\cite{Gabadadze:2006xm}
% \bibitem{Gabadadze:2006xm}
  G.~Gabadadze,
  ``A model for cosmic self-acceleration,''
  arXiv:hep-th/0612213.
  %%CITATION = HEP-TH/0612213;%%


%%%%%%%%%%%%%%%%%%%%%%%%%%%%%%%%%%%%%%%%%%%%%%%%%%%%%%%%%
% f(R) gravity


%\cite{Copeland:2006wr}
\bibitem{Copeland:2006wr}
  E.~J.~Copeland, M.~Sami and S.~Tsujikawa,
  ``Dynamics of dark energy,''
  Int.\ J.\ Mod.\ Phys.\  D {\bf 15}, 1753 (2006)
  [arXiv:hep-th/0603057].
  %%CITATION = IMPAE,D15,1753;%%



%%%%%%%%%%%%%%%%%%%%%%%%%%%%%%%%%%%%%%%%%%%%%%%%%%%%%%%%%
% 6d-Sled Higgs coupling



%\cite{Beauchemin:2004zi}
\bibitem{Beauchemin:2004zi}
  P.~H.~Beauchemin, G.~Azuelos and C.~P.~Burgess,
  ``Dimensionless coupling of bulk scalars at the LHC,''
  J.\ Phys.\ G {\bf 30}, N17 (2004)
  [arXiv:hep-ph/0407196];
  %%CITATION = JPHGB,G30,N17;%%
J.~J.~van der Bij and S.~Dilcher,
  ``A higher dimensional explanation of the excess of Higgs-like events at CERN
  LEP,''
  Phys.\ Lett.\  B {\bf 638}, 234 (2006)
  [arXiv:hep-ph/0605008].
  %%CITATION = PHLTA,B638,234;%%




\end{thebibliography}
\end{document}